\documentclass[%
 preprint,
 longbibliography,
nofootinbib,
amsmath,
amssymb,
aps,
]{revtex4-1}

\usepackage{graphicx}
\usepackage{dcolumn}
\usepackage{bm}
\usepackage{todonotes}
\usepackage{subfig}
\usepackage{fullpage}

\usepackage{amsmath}
\usepackage{hyperref}
\usepackage{cleveref}
\usepackage{amsfonts}
\usepackage{amssymb}
\usepackage{algorithm}
\usepackage{algpseudocode}
\usepackage{float}
\usepackage{relsize}
\usepackage{epstopdf}
\usepackage{color}
\usepackage[mathscr]{eucal}

\usepackage[utf8]{inputenc}
\usepackage[english]{babel}
\usepackage{amsthm}

\newcommand{\nc}{\newcommand}
\nc\be{\begin{eqnarray}}
\nc\ee{\end{eqnarray}}
\nc\bs{\begin{eqnarray*}}
\nc\es{\end{eqnarray*}}
\nc\qu{\quad}
\nc\tends{\rightarrow}
\nc\dint{\int\!\int}

\newcommand{\bes} {\begin{eqnarray*}}
\newcommand{\ees} {\end{eqnarray*}}
\newcommand{\dd} \partial

\newcommand{\non} \nonumber

\nc\cA{{\cal{A}}}
\nc\cB{{\cal{B}}}
\nc\cZ{{\cal{Z}}}
\nc\cX{{\cal{X}}}
\nc\cS{{\cal{S}}}
\nc\cR{{\cal{R}}}
\nc\cW{{\cal W}}
\nc\cU{{\cal U}}
\nc\cP{{\cal P}}
\nc\cF{{\cal{F}}}
\nc\cG{{\cal{G}}}
\nc\bz{\bar{z}}
\nc\bZ{\bar{Z}}
\nc\baz{\bar{\zeta}}
\nc\bw{\bar{w}}
\nc\bW{\bar{W}}
\nc\bX{\bar{X}}
\nc\bet{\bar{\eta}}
\nc\bp{\bar{\phi}}
\nc\bP{\bar{\Phi}}
\nc\bc{\bar{\chi}}
\nc\baf{\bar{f}}
\nc\baF{\bar{F}}
\nc\ov{\overline}
\nc\oW{\ov{W}}

\nc\hA{\hat{A}}
\nc\hB{\hat{B}}
\nc\ho{\hat{O}}
\nc\hG{\hat{\Gamma}}
\nc\hS{\hat{S}}
\nc\hO{\hat{\Omega}}
\nc\bO{\ov{\Omega}}
\nc\gham{\hat{\gamma}}
\nc\gcam{\check{\gamma}}
\nc\z{\zeta}
\nc\s{\sigma}
\nc\ep{\epsilon}
\nc\up{\upsilon}
\nc\lam{\lambda}
\nc\Lam{\Lambda}
\nc\sig{\sigma}
\nc\om{\omega}
\nc\kap{\kappa}
\nc\gam{\gamma}
\nc\Om{\Omega}
\nc\pa{\partial}
\nc\dom{\pa\Omega}
\nc\domt{\pa\tilde{\Omega}}
\nc\doh{\pa\hat{\Omega}}
\nc\pad[2]{\frac{\pa #1}{\pa #2}}
\nc\padd[2]{\frac{\pa^2 #1}{\pa {#2}^2}}
\nc\pard[3]{\frac{\pa^2 #1}{\pa {#2}\pa {#3}}}
\nc\nd[2]{\frac{d #1}{d #2}}
\nc\ndd[2]{\frac{d^2 #1}{d {#2}^2}}
\nc\ds{\displaystyle}
\nc\del{\nabla}
\nc\lap{\nabla^2}
\nc\ud{|\z|\leq 1}

\nc\capil{\mbox{Ca}}

\nc\pt{\tilde{a}}
\nc\ft{\tilde{f}}
\nc\vt{\tilde{v}}
\nc\wt{\tilde{w}}
\nc\nt{\tilde{\nu}}
\nc\xit{\tilde{\xi}}
\nc\xt{\tilde{x}}
\nc\etat{\tilde{\eta}}
\nc\nht{\tilde{\hat{\eta}}}
\nc\taut{\tilde{\tau}}
\nc\Taut{T}
\nc\zt{\tilde{\z}}
\nc\nut{\tilde{\nu}}

\nc\At{\tilde{\cA}}
\nc\qt{\tilde{B}}
\nc\Ct{\tilde{C}}
\nc\dt{\tilde{D}}
\nc\Dt{\tilde{\cU}}
\nc\Et{\tilde{L}}
\nc\Ft{\tilde{F}}
\nc\Ht{\tilde{H}}
\nc\Kt{\tilde{K}}
\nc\Lt{\tilde{K}}
\nc\Mt{\tilde{M}}
\nc\Nt{\tilde{N}}
\nc\Pt{\tilde{\Phi}}
\nc\Qt{\tilde{Q}}
\nc\Rt{\tilde{R}}
\nc\St{\tilde{\cS}}
\nc\Tt{\tilde{\tau}}
\nc\Wt{\tilde{W}}
\nc\cWt{\tilde{\cW}}
\nc\Xt{\tilde{X}}

\nc\vs{\varsigma}
\nc\vp{\varpi}
\nc\ve{\varepsilon}
\nc\vepa{\ve_{\parallel}}
\nc\vepe{\ve_{\perp}}

\begin{document}

\preprint{APS/123-QED}

\title{Modeling and design optimization for pleated membrane filters}

\author{Yixuan Sun}
\email{ys379@njit.edu}
\affiliation{Department of Mathematical Sciences and Center for Applied Mathematics and Statistics, New Jersey Institute of Technology, Newark, NJ 07102-1982, USA}

\author{Pejman Sanaei}
\affiliation{
Department of Mathematics, New York Institute of Technology, New York, NY 10023-7692, USA}
\author{Lou Kondic, Linda J. Cummings}
\affiliation{
Department of Mathematical Sciences and Center for Applied Mathematics and Statistics, New Jersey Institute of Technology, Newark, NJ 07102-1982, USA}%

\begin{abstract}
Pleated membrane filters, which offer larger surface area to volume ratios than unpleated membrane filters, are used in a wide variety of applications. However, the performance of the pleated filter, as characterized by a flux-throughput plot, indicates that the equivalent unpleated filter provides better performance under the same pressure drop. Earlier work (Sanaei~\& Cummings 2016) used a highly-simplified membrane model to investigate how the pleating effect and membrane geometry affect this performance differential. In this work, we extend this line of investigation and use asymptotic methods to couple an outer problem for the flow within the pleated structure to an inner problem that accounts for the pore structure within the membrane. We use our new model to formulate and address questions of optimal membrane design for a given filtration application. 
\end{abstract}

\maketitle

\section[short]{Introduction\label{2intro}}

At the most basic level, membrane filtration is a process of separation, whereby undesired/desired particles are removed from a fluid suspension (known as a {\it feed}) by passing through a porous membrane. 
Membrane filters are commonly used in many applications and feature in many aspects of daily life, such as drinking water production \citep{hoslett2018}, beer purification \citep{sman2012}, vaccine purification \citep{emami2018}, food and dietary supplement production \citep{yogarathinam2018}, natural gas purification \citep{liu2018} and many more. With growing interest from both industrial and academic sectors in understanding the various types of filtration processes in use, and improving filter performance, the past few decades have seen many publications in this area, including excellent review articles that cover various aspects of membrane filtration (see, for example, \cite{tang2011, iritani2013, reis2007, ulbricht2006}). Nonetheless, with the majority of publications on the experimental side, simple and readily-applicable ``first principles'' mathematical models, that can explain and predict membrane filter performance, and guide improvements to membrane filter design, are still lacking. 

Since the function of the membrane filter is to remove particles from the feed suspension, fouling of the membrane is an unavoidable part of successful filtration, and considerable research effort has been devoted to understanding the fouling mechanisms (see, for example, \cite{bowen1995a, ho2000,orsello2006, bolton2006a,rebai2010,giglia2012,iliev2018,lohaus2018}, among many others), with a view to elongating the useful life-span of a filter. Four basic fouling mechanisms have been identified in the literature (see, e.g. \cite{grace1956, hermia1982}), often characterized as follows: (i) {\it standard} or {\it adsorptive blocking} (particles smaller than membrane pores enter and deposit on the wall to shrink the pores);(ii) {\it complete blocking} (particles larger than pores deposit at a pore entrance on the membrane surface and block the pore); (iii) {\it  intermediate blocking} (as for complete blocking except that pores are not completely sealed); and (iv) {\it  cake filtration} (once the pores on the membrane surface are blocked, further particles stack up on the membrane surface, forming a ``cake layer'').

In practice, it is rare that an entire filtration process is described well by a single fouling mechanism due to the complex composition of the feed and convoluted interaction between the feed and the porous medium (the {\it filter}). Often data indicate that two or more mechanisms are in operation simultaneously or sequentially ({\it e.g.}~\cite{tracey1994}), and several authors have proposed models to account for multiple fouling modes. For example, \cite{ho2000}, \cite{bolton2006a}, and \cite{sanaei2016} each proposed different models to account for two distinct fouling mechanisms; and \cite{orsello2006} published a model to account for three sequential fouling mechanisms. In this last work, the pore is first constricted by small particles (standard blocking dominates); then, once the pore is sufficiently small, further particles are sieved out (complete blocking dominates); and finally a cake layer is formed that dominates the end stage of the filtration process. For an appropriate choice of model parameters, a good agreement with experimental results was obtained.

Although inclusion of more fouling mechanisms may give a more complete picture, this comes at the expense of a larger number of model parameters that must be determined and explored. Moreover, it is often the case that a single fouling mode can dominate the majority of the filtration process, for a sufficiently simple feed. For example, \cite{beuscher2010} 
developed a single-mode fouling model considering only complete blocking within a membrane modeled as a layered network of pores, connected at layer junctions. Pore sizes in each layer are drawn from a probability distribution, and particle transport through the pore network also follows a specified
probabilistic model, with certain (physically-motivated) restrictions. This model shows reasonable agreement with an experimental dataset. Though a very different model compared to the type we shall derive and study, it yields results that have certain features in common with ours, and we will return to this work later.

\begin{figure}
\centering
{\scriptsize(a)}\includegraphics[scale=0.5]{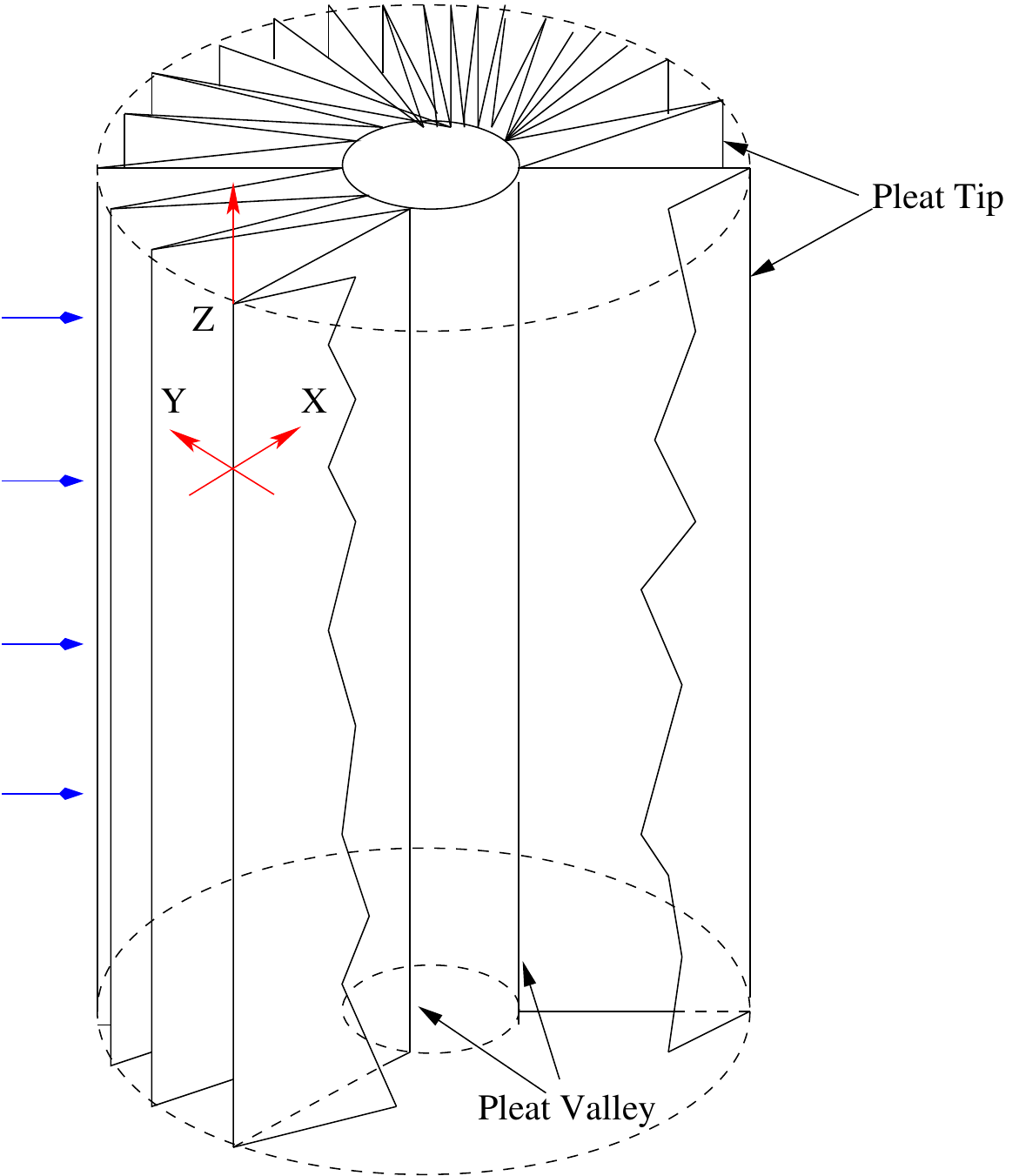}
{\scriptsize(b)}\includegraphics[scale=0.43]{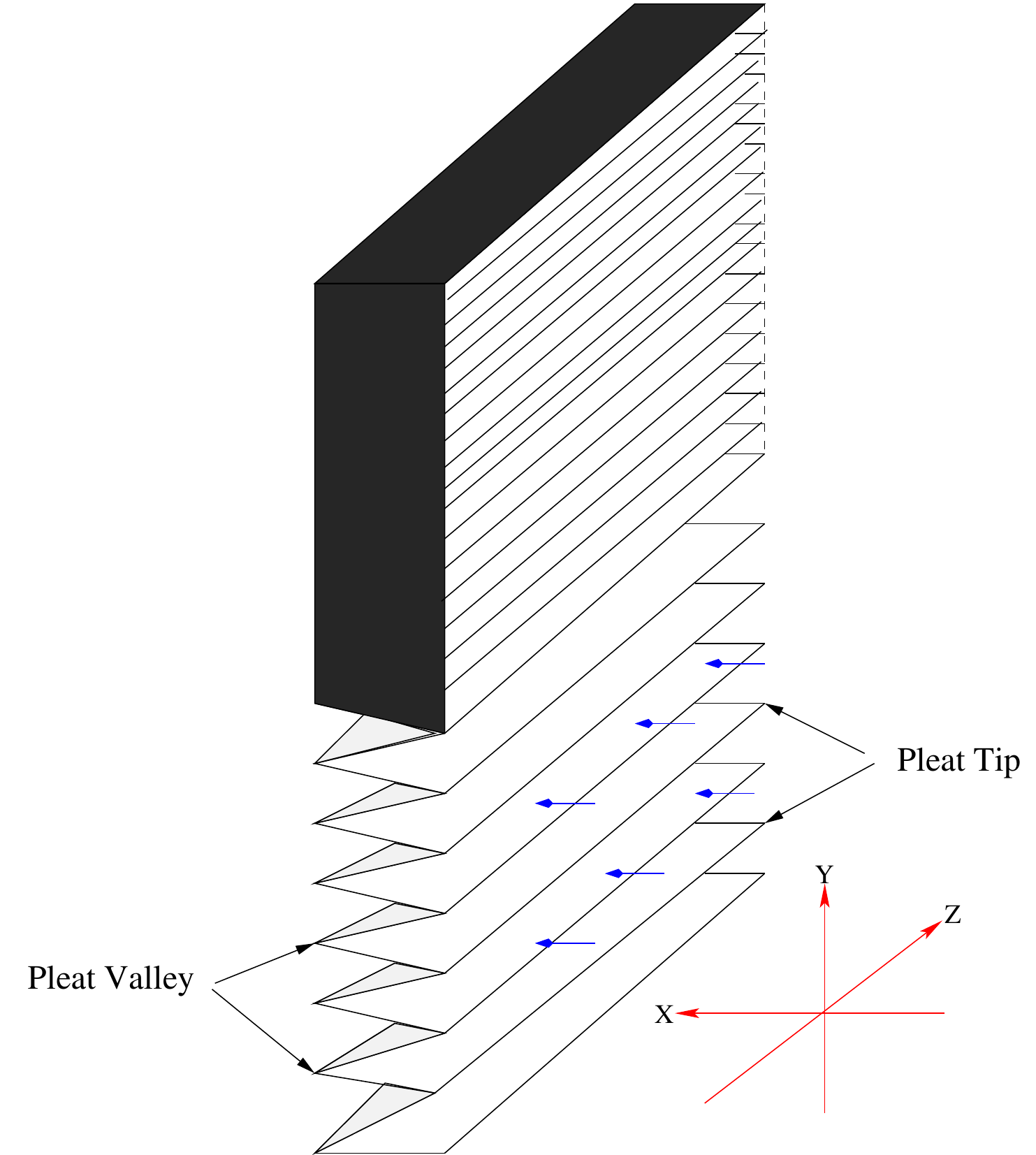}
\caption{\footnotesize{Sketches of (a) a cylindrical pleated membrane filter cartridge; (b) a rectangular pleated membrane filter cartridge. Blue arrows indicate the flow direction. }
}
\label{2real-cartridge}
\end{figure}

Membrane filter design can vary widely depending on the application, in terms of both internal pore structure (microscale design), and how the membrane is deployed (macroscale design). In many applications it is desirable to have a large membrane area available to maximize throughput, while simultaneously keeping the volume of the filtration unit to a minimum. {\it Pleated} membrane filters are commonly used to achieve this tradeoff, and these are the type of filters that we consider in this paper (though many aspects of our modeling are more generally applicable). 
Pleated filters may have different geometries: some are cylindrical (see figure \ref{2real-cartridge}(a)); while some are rectangular (figure \ref{2real-cartridge}(b)). In most cases the membrane is sandwiched between two supporting layers 
 and the resulting three-layer structure is folded (pleating) and fixed (e.g. via heating, \cite{brown2011}) to give permanence to the shape and structure of the pleated filter. 
Several key design factors used in characterization of pleated filters have been identified in the literature, and a 
detailed description can be found in, {\it e.g.}, \cite{jornitz2006}, \cite{reis2007} and \cite{brown2011}. In our study, we focus on pleated filters with high {\it pleat packing density} (PPD), in which it is assumed that air gaps between adjacent pleats are negligible. Earlier work \citep{sanaei2016} used a highly-simplified membrane model to investigate how the pleating effect and membrane geometry affect the performance. In this work, we extend this line of investigation by using asymptotic methods to couple an outer problem for the flow within the pleated structure to a detailed inner problem that accounts for the pore structure (shape) within the membrane, and also by incorporating a model for the transport of small particles that lead to adsorptive fouling.

 Although design improvements can be arrived at by trial and error, making prototypes and testing them can be costly and the process can be hard or impossible to systematize \citep{ulbricht2006}.  With this in mind, we are motivated to set up a simplified mathematical model to describe the flow through and fouling of pleated membrane filters, which can be used to evaluate filtration performance and ultimately to optimize the design. In this study, we focus on the effect that the membrane's microstructure has on the filtration performance; in particular, the effects of varying pore size in the depth of the filter membrane.  
In order to keep the number of adjustable parameters to a minimum and to obtain clearer predictions for the impact of membrane design on filtration performance, we choose to focus on a single fouling mode: standard (or adsorptive) blocking. In \S~2, we present the mathematical model; in \S~3, we formulate the optimization problem for the microscale membrane design and present results; and in \S~4, we summarize our conclusions and discuss some ideas for future investigation. 

\begin{figure}
\centering
{\scriptsize (a)}{\includegraphics[scale=.45]{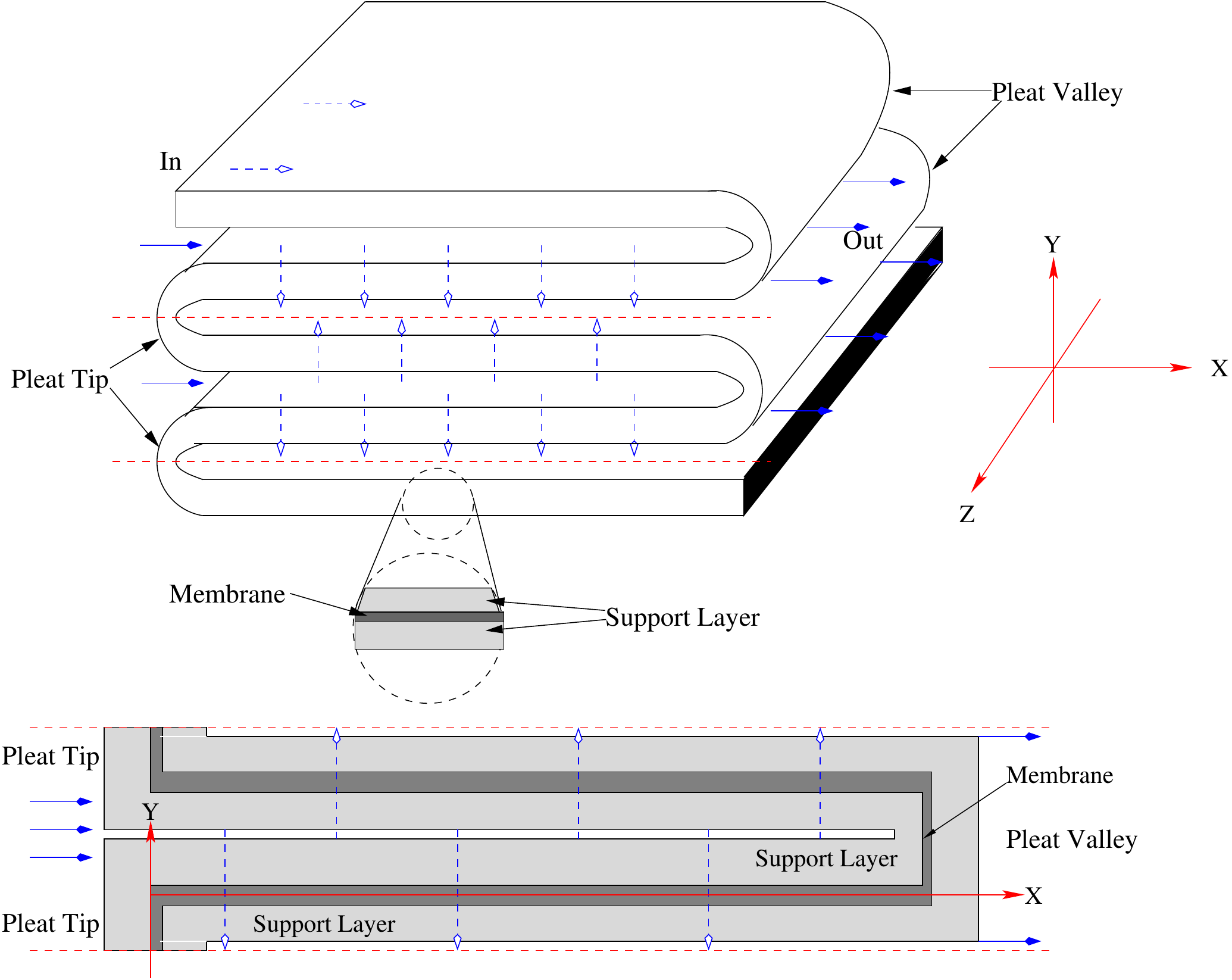}}\\
\vspace{0.1in}
{\scriptsize (b)}\includegraphics[scale=0.45]{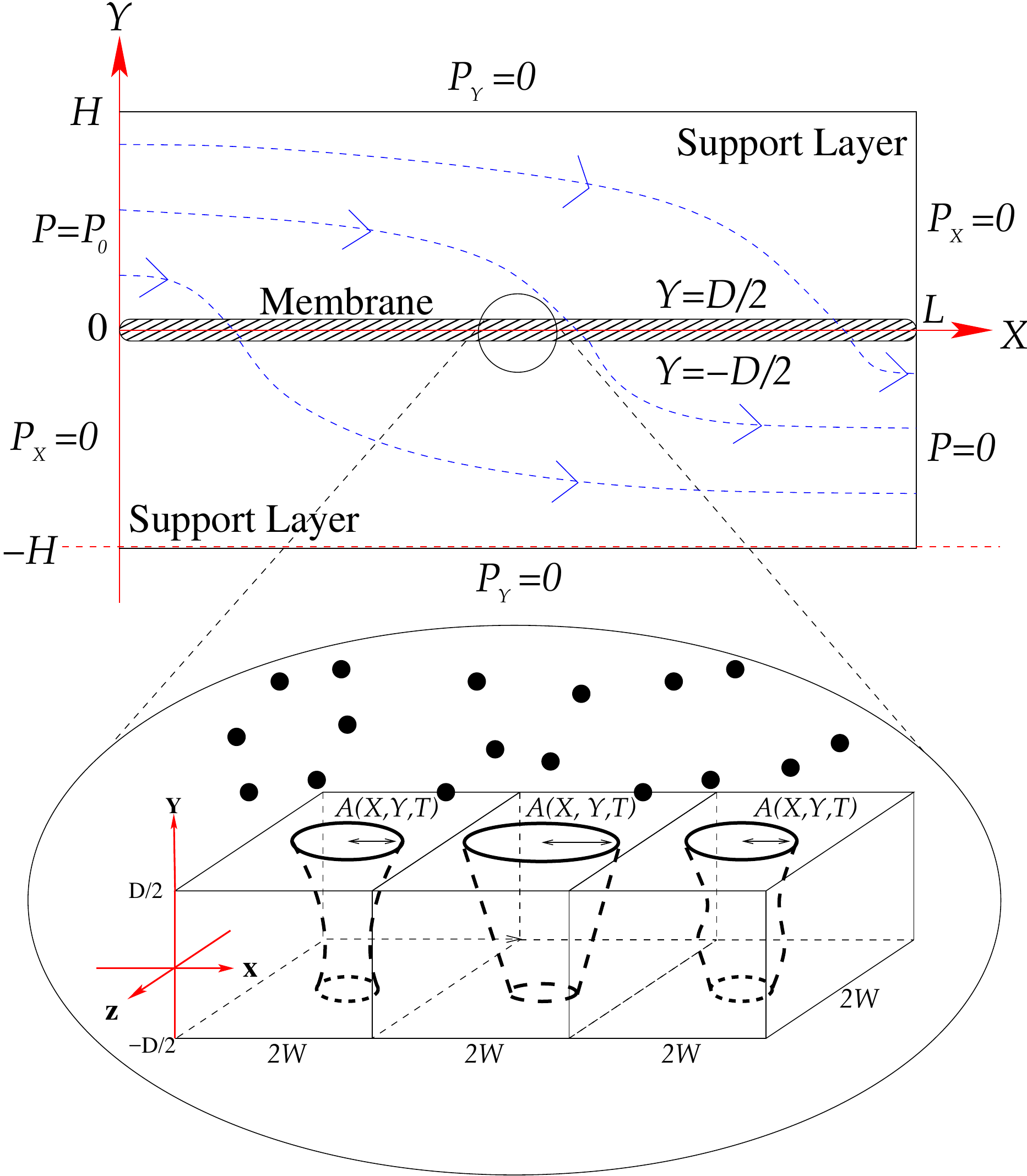}
\caption{\footnotesize{(a) Upper: Schematic, showing a few pleats. The region between the red dashed lines indicates a single complete pleat, assumed to repeat periodically. The zoom-in indicates the three-layer structure, with pale grey denoting the support layers and dark grey the membrane layer (in reality much thinner than the support layers). Blue arrows indicate the flow direction. Lower: A single pleat period, indicating how the geometry is idealized in the model, with the same color coding as the zoom-in. 
  (b)  Simplified domain (half the pleat) showing boundary conditions at inlet and outlet and schematic streamlines.  Symmetry is assumed about $Y=\pm H$, and support layer plus membrane occupies the whole space (no air gap). The zoom-in shows the membrane pore morphology.}}
\label{2idealized}
\end{figure}

\section{Mathematical modeling\label{2fluid}}

\subsection{Modeling assumptions: outline\label{2assumptions}}

In this study, we will focus ostensibly on the rectangular pleated membrane filter, though if one neglects the curvature of the cylindrical filter cartridge then our results are quite generally applicable. As noted in the Introduction, we consider pleated filters with high pleat packing density (PPD), so that we may assume that all flow regions are occupied by either porous support material or filter membrane, with no air gaps. The {\it pleat tip} is defined to be the membrane fold on the inflow side, while the {\it pleat valley} is the fold on the outflow side. Figure \ref{2idealized} indicates how we simplify the geometry, first by neglecting dependence on the coordinate $Z$ that runs parallel to the pleat tips/valleys; and then (if considering the cylindrical cartridge) neglecting the curvature of the cylinder. This reduces the problem to 2D in the $(X,Y)$-plane, where $X$ is the direction along the pleat (from tip to valley), and $Y$ is perpendicular to the membrane in each pleat (the azimuthal direction in the cylindrical cartridge). We further simplify by neglecting the curvature at the pleat tip and valley, which allows us to view each pleat as one of a periodic array of rectangles, indicated in the lower part of figure \ref{2idealized}(a). Symmetry of each rectangle (pleat) about the centerline ($Y=H$) is assumed, which gives our simplified half-pleat domain, indicated in figure \ref{2idealized}(b).

We assume that Darcy's law governs flow in both the support layers and the filter membrane. Both are porous media, though the support material has a much higher permeability ($K$) than that of the membrane ($K_{\rm m}$): $K\gg K_{\rm m}$; this will be important in our asymptotic analysis later. We further assume a no-flux condition at pleat tips and valleys, justified on the basis that both the support material and the membrane there will be tightly compressed if PPD is high, so that these regions will have much higher resistance than the main length of the pleat. Support permeability $K$ is assumed to be constant in both space and time, reflecting an assumption that fouling of support layers (which are not designed to capture particles) does not occur: the pores of these layers are much larger than those of the membrane filter. In general, however, $K_{\rm m}$ will vary in both time and space due to fouling: $K_{\rm m}(X,T)$. The fouling occurs on a time scale much longer than the membrane transit time for the fluid, justifying an assumption that the flow problem may be solved quasi-statically for the instantaneous membrane permeability $K_{\rm m}(X, T)$. As the fouling progresses the pore radii shrink and $K_{\rm m}$ decreases (as explained below), which will in turn change the pressure and velocity fields within the system.  

The membrane has thickness $D$, the supporting layers have thickness $H-D/2$, and the length from the pleat tip to pleat valley is $L$ (also referred to as pleat depth in the literature). The membrane occupies the region $0\leq X\leq L$, $-D/2\leq Y\leq D/2$, and feed passes through the pores in the negative $Y$-direction. In pleated filters used in applications \citep{ulbricht2006}, the pleat depth is much larger than the membrane and support layer thicknesses; and the membrane thickness is often much smaller than the thickness of the support layer, i.e. $L\gg H \gg D$. We assume that each pore traverses the membrane from upstream to downstream side without branching, is contained within a square prism of dimensions $2W\times 2W \times D$, and that the membrane consists of a periodic array of such prisms (see the zoom-in figure \ref{2idealized}(b)). Pores are modeled as slender tubes of circular cross-section, whose axis coincides with that of the containing prism. The pore radius is denoted by $A(X,Y,T)$, allowing for variation in both the plane ($X$) and the depth ($Y$) of the membrane, as well as time ($T$) due to the fouling. This description is entirely appropriate for ``track-etched'' membranes (in which pores are made by etching the nuclear tracks left by radiation \cite{apel2001}), but may also be considered as a reasonable model for membranes with more general pore structure in a depth-averaged sense.  We also assume the pore-containing prism is long and thin, i.e. $W\ll D$.
 The lengthscale separations $W \ll D \ll L$ justify the slender pore assumption (with slowly-varying radius in $Y$), as well as our assumption (implicit in the above) that pores are sufficiently numerous in the $X$ direction that they may be considered as continuously distributed in $X$, with radius $A(X,Y,T)$ also continuously varying in $X$.

We assume the feed is a dilute suspension, which we treat as an incompressible Newtonian fluid. We denote the concentration of particles (assumed identical and much smaller than pore radii) in the fluid by $C$, with $C\equiv C_0$ in the inflowing feed solution. Consistent with the assumption of no fouling in the support layers, we take
$C=C_0$ in the whole region $[0,L]\times [D/2,H]$; $C$ will vary in both space and time within the membrane as adsorptive fouling occurs, as described below. 

\subsection{Governing equations\label{2model}}

We study the case where flow through the pleated filter unit is driven by a constant pressure drop $P_0$, with no-flux conditions at pleat tips and valleys. With the further assumptions of periodicity and symmetry within each pleat 
(indicated in figure \ref{2idealized}(a) and outlined in \S~\ref{2assumptions}), we can simplify the problem domain to be half of the pleat (figure \ref{2idealized}(b)), with no flux (symmetry) boundary conditions imposed at $Y=\pm H$.
Based on the assumptions set out in \S~2.1, we model the flow in both the supporting and the membrane layers with Darcy's law. The velocity $\mathbf{U}=(U,V)$ within the support layers is then given in terms of the pressure $P(X,Y)$ by
\begin{equation}
  \mathbf{U}=(U,V)= -\frac{K}{\mu} \nabla P, \quad \nabla = (\partial_X,\partial_Y), \quad 0\leq X\leq L,\, D/2\leq |Y| \leq H.
  \label{2darcy}
\end{equation}
Incompressibility of the feed solution requires
\begin{equation}
\nabla\cdot \mathbf{U}=0 \quad\Rightarrow\quad
\nabla^2 P=0,  \quad 0\leq X\leq L,\, D/2\leq |Y| \leq H,
\label{2incomp}
\end{equation}
within the support layers, under the stated assumption that support permeability $K$ does not vary spatially. We have the following boundary conditions on the pressure within the support layers: 
\begin{equation}
P^+(0,Y)=P_0, \, P^+_X (L,Y)=0, \, P^+_Y(X,H)=0, 
\label{2BC+}
\end{equation}
\begin{equation} 
P^-_X(0,Y)=0, \, P^-(L,Y)=0, \, P^-_Y(X,-H)=0,
\label{2BC-}
\end{equation}
where we use $\pm$ superscripts to distinguish between quantities evaluated for $Y\gtrless 0$ respectively, on either side of the membrane. 

As outlined in \S~2.1, we model membrane pores as a distribution of slender tubes with circular cross-section spanning the membrane, of length $D$ and radius $A(X,Y,T)$. For our preliminary investigations, we assume that initially all pores are identical (homogeneity in the plane of the membrane), with radius varying only in the $Y$ direction, $A(X,Y,0)=A_0(Y)$ (note, however, that with the $X$-dependent pressure distribution and geometry, fouling will vary with respect to $X$ and thus the local pore radius $A$ will depend on $X$ for $T>0$). With this membrane structure, it is reasonable to assume that the Darcy flow through the membrane is approximately unidirectional, $\mathbf{U}_{\rm m} =(0,V_{\rm m})$.\footnote{For membranes that are not of this simple ``track-etched'' type, our additional lengthscale assumption $D\ll H \ll L$ helps justify this unidirectional flow approximation.} Incompressibility then gives $\partial V_{\rm m}/\partial Y =0$, with 
\begin{equation}
V_{\rm m}(X,T) = -\frac{K_{\rm m}}{\mu} \frac{\partial P_{\rm m}}{\partial Y}, \quad -D/2\leq Y\leq D/2.
\label{2darcy_pore}
\end{equation} 
Since the pressure gradient $\partial P_{\rm m}/\partial Y$ is independent of $Y$, this equation may be rewritten as 
\be
|V_{\rm m}| =\frac{K_{\rm m}}{\mu D} \left[ \left. P^+ \right|_{Y=D/2}- \left. P^- \right|_{Y=-D/2} \right], \qquad 0\leq X\leq L,
\label{2flux-cty0}
\ee
where continuity of the pressure between support layer and membrane at the membrane boundaries was used. Here, the membrane permeability $ K_{\rm m} $ is related to the local pore radius $A$ by
\begin{equation}
{ K }_{\rm m} (X,Y,T)=\frac { \pi { A^4(X,Y,T) } }{ 32{ W }^{ 2 } },
\label{2Km}
\end{equation}
which follows from the Hagen-Poiseuille formula, see, {\it e.g.}, \cite{probstein1994} 
(recall that $2W$ is the size of the pore-containing period box). 
Continuity of flux between the support layers and membrane gives 
\be
|V_{\rm m}|= \frac{K}{\mu} \left. \frac{\partial P^+}{\partial Y} \right|_{Y=D/2}=  \frac{K}{\mu} \left.\frac{\partial P^-}{\partial Y} \right|_{Y= -D/2} ,
\qquad 0\leq X\leq L.
\label{2flux-cty}
\ee
The (cross-sectionally averaged) velocity of the fluid within the pore, $V_{\rm p}(X,Y,T)$, is related to the superficial Darcy velocity within the membrane, $V_{\rm m}(X,T)$, by 
\begin{equation}
4W^2 V_{\rm m} = \pi A^2 V_{\rm p}.
\label{2porevel-superficialvel}
\end{equation}
The local pore radius $A$ (and hence the membrane permeability at that location) changes in time due to the adsorptive fouling, modeled by accounting for particle transport and deposition within pores. Following earlier work \citep{sanaei2017}, we propose a simple advection and deposition model for the particle concentration $C(X,Y,T)$ within pores, 
\begin{equation}
V_{\rm p} \frac{\partial C}{\partial Y}=-\Lambda \frac{C}{A},\quad C(X,\frac{D}{2} ,T)=C_0.
\label{2deposition}
\end{equation}
This model comes from an asymptotic analysis of the full advection and diffusion equation for a suspension of small particles passing through a slender tube; the detailed justification can be found in \citet{sanaei2017}. 
The dimensional constant $\Lambda$ measures the strength of attraction between the particles and the pore wall that gives rise to the deposition. The pore radius shrinks in response to the deposition: consistent with (\ref{2deposition}) we propose
\begin{equation}
\frac{\partial A(X,Y,T)}{\partial T}= -\Lambda \alpha C(X,Y,T), \quad A(X,Y,0)=A_0 (Y),
\label{shrinkage}
\end{equation}
for some constant $\alpha$ (proportional to the particle volume).
This formulation with $A_0(Y)$ assumes that all pores in the membrane are identical initially. We will later briefly consider initial pore profiles that can vary in the $X$-direction also, so as to allow investigation of the possible effects of nonuniform pore size distribution in the plane of the membrane.

\subsection{ Nondimensionalization \& Asymptotic Analysis\label{4sec:press}}

\begin{table}
\centering
\begin{tabular}{|l|l|l|}
\hline
{\bf Parameter} & {\bf Description} & {\bf Typical value}\\
\hline
$L$ &\mbox{Length of the pleat} & $1.3$ \mbox{cm}\\
$H$ & \mbox{Support layer thickness}&$1$ \mbox{mm} \\
$D$ &\mbox{Membrane thickness}& $300$ \mbox{$\mu$m}\\
$W$ &\mbox{Pore prism lateral dimension}& 2~$\mu$m (very variable)\\
$P_0$ & \mbox{Pressure drop} & Depends on application\\
$K$ & \mbox{Support layer permeability} & \mbox{4$\times$10$^{-11}$~m$^2$} (very variable)\\
$K_{\rm m0}$ & \mbox{Representative membrane permeability} & 4$\times$\mbox{10$^{-13}$~m$^2$} (very variable)\\
\hline
\end{tabular}
\caption{\footnotesize{Approximate dimensional parameter values~\citep{kumar2014}. Based on the application, the pore size may vary from 1 nm to 10 $\mu$m \citep{reis2007}. } 
}\label{2t:parameters1}.
\label{2params}
\end{table}

\begin{table}
\centering
\begin{tabular}{|l|l|l|}
\hline
{\bf Parameter} & {\bf Formula} & {\bf Typical value}\\
\hline
$\ep $ & $H/L$ & $0.077$\\
$\delta$ & $D/H$& $0.3$\\
$\Gamma$& $K_{\rm m0}/(K  \epsilon^2 \delta)$ & 5.53\\
\hline
\end{tabular}
\caption{\footnotesize{Approximate dimensionless parameter values (from Table~\ref{2t:parameters1}). 
}}
\label{2t:parameters2}
\end{table}

\subsubsection{Nondimensionalization}

In order to identify and exploit asymptotic simplifications, we introduce the following scales and dimensionless variables. Consideration of the support layers suggests the scales
\be
(X,Y)=(L x,H y),  \quad P^{\pm}(X,Y,T)=P_0 p^{\pm}(x,y,t),
\label{support-scalings}
\ee
while consideration of the membrane filter leads to the remaining scalings, 
\be
Y= D\tilde { y }, & \quad A(X,Y,T)=W a(x,\tilde { y } ,t), \label{2scalings1}\\
K_{\rm m}(X,Y,T)=K_{\rm m0} k_{\rm m}(x,\tilde{y},t), & \quad P_{\rm m}(X,Y,T)=P_0 p_{\rm m}(x,\tilde{y},t), 
\label{2scalings2} \\
 { V_{\rm m} (X,T)=\frac { { K }_{\rm m0 }{ P }_{ 0 } }{ \mu D } { v }_{\rm m } }(x, t), & \quad C(X,Y,T)={ C }_{ 0 }c(x,\tilde { y } ,t). 
\label{2scalings3}
 \ee
Time is scaled on the pore shrinkage timescale,
\be
 \quad T=\frac{W}{\Lambda\alpha C_0}t.
 \label{2scalings4}
 \ee
Note that we have introduced two scaled coordinates in the $Y$-direction: $y$ in (\ref{support-scalings}) and $\tilde{y}$ in (\ref{2scalings1}), relevant to the support layer and the membrane, respectively. The chosen scalings lead naturally to two dimensionless length ratios: $\epsilon=H/L$, the aspect ratio of the pleat; and $\delta=D/H$, the ratio of membrane thickness to support layer thickness. Both will be assumed small in the following: $\epsilon\ll 1$, $\delta\ll 1$ (no further assumption is needed on the relative sizes of $\epsilon$ and $\delta$). In (\ref{2scalings3}), $K_{{\rm m} 0}$ is a representative value for the initial membrane permeability. For simplicity, we choose $ { K }_{\rm m0 }= { \pi { W }^{ 2 } }/{ 32 }  $, then equation (\ref{2Km}) gives the dimensionless membrane permeability as
\be
 { k }_{ \rm m }(x,\tilde{y},t)={ a^4(x,\tilde{y},t) }.
\label{2km}
\ee
\subsubsection{Darcy flow in supporting layers}

The dimensionless governing equations and boundary conditions in the supporting layers $ {\delta}/{2} \leq |y| \leq 1$ are
\be
\ep^2 p^+_{xx} + p^+_{yy}=0, \quad \delta/2 \leq y \leq 1, \label{p+} \\
p^+ (0,y)=1, \quad p^+_x (1,y)=0, \quad p^+_y (x,1)=0, \label{p+bc} \\
\ep^2 p^-_{xx} + p^-_{yy}=0, \quad -1\leq y \leq -\delta/2, \label{p-} \\
p^-_x (0,y)=0, \quad p^- (1,y)=0, \quad p^-_y (x,-1)=0, \label{p-bc}
\ee
where $\ep$ and $\delta$ are defined in Table \ref{2t:parameters2}, and we have suppressed the time dependence with the understanding that the time taken for significant fouling of the membrane to occur is much longer than that taken for fluid to transit the filter unit (note that with this assumption, the fouling is the only unsteady process in our model).
This system is closed by enforcing flux continuity across the membrane, equations (\ref{2flux-cty0}) and (\ref{2flux-cty}), giving 
\be
 p^+_y |_{y=\delta/2} =  p^-_y  |_{y=-\delta/2} =- \Gamma v_{\rm m},
\label{2flux-cty-dimless}
\ee
where the dimensionless parameter $\Gamma$, defined by 
\be
\Gamma=\frac{K_{\rm m0}}{K  \epsilon^2 \delta} \label{gamma},
\ee
captures the relative importance of the resistance (inversely proportional to the permeability) of the supporting material to that of the membrane, such that if $\Gamma \gg 1$ the supporting material provides most of the resistance whereas if $\Gamma \ll 1$ the membrane provides most of the resistance. In our analysis we assume $\Gamma$ is O(1) (see Table \ref{2t:parameters2} for a value for typical pleated filters).

\subsubsection{Flow and fouling within the membrane}
The dimensionless equation for Darcy flow through the membrane layer, defined by $-1/2 \leq \tilde{y}\leq 1/2$, is:
\be
v_{\rm m}=-k_{\rm m}\frac{\partial p_{\rm m}}{\dd \tilde{y}}=-a^4\frac{\dd p_{\rm m}}{\dd \tilde{y}}.
\label{vm_nd}
\ee
With the incompressibility condition ${\partial v_{\rm m}}/{\partial \tilde{y}}=0$, and the continuity of pressure at the membrane surfaces, i.e. $p_{\rm m}(x,1/2)=p^+(x,{\delta}/{2})$ and  $p_{\rm m}(x,-1/2)=p^-(x,-{\delta}/{2})$, we have
\be
v_{\rm m}=\frac{p^-(x,-\frac{\delta}{2})-p^+(x,\frac{\delta}{2})}{\int_{ -\frac{1}{2} }^{\frac{1}{2}  }{ \frac{d \tilde{y}}{a^4} } },
\label{vm_int}
\ee
with $p^-, p^+$ determined by the support layer model outlined above. 
Note here the use of the two different length scales, in the supporting layer ($y=Y/H$; numerator of (\ref{vm_int})) and within the membrane layer ($\tilde{y}=Y/D$; denominator of (\ref{vm_int})).  
Equation (\ref{2deposition}) for the particle concentration $c(x,\tilde{y},t)$ within the membrane becomes
\begin{equation}
v_{\rm m} \frac{\partial c}{\partial \tilde{y}}=-\lambda ac,\quad c(x,\frac{1}{2} ,t)=1,
\label{2deposition_nd}
\end{equation}
where the dimensionless deposition coefficient $\lambda$ is given by
\be
\lambda=\frac{\pi \mu D^2 \Lambda}{4 W K_{\rm m0} P_0}.
\label{lambda}
\ee
The pore radius evolution equation (\ref{shrinkage}) becomes
\begin{equation}
\frac{\partial a(x,\tilde{y},t)}{\partial t}= - c(x,\tilde{y},t), \quad a(x,\tilde{y},0)=a_0 (\tilde{y}).
\label{2shrinkage_nd}
\end{equation}

\subsubsection{Asymptotic solution}

We seek asymptotic solutions for $p^{\pm}$ in the distinguished limit $\Gamma=O(1)$, $\ep \ll 1$ by expanding $p^{\pm}$ in powers of $\ep$ as follows:
\be
p^+(x,y)= p_0^+(x,y) +\ep^2  p_1^+(x,y)+\cdots, 
\label{p+sol} \quad
p^-(x,y)= p_0^-(x,y) +\ep^2 p_1^-(x,y) +\cdots,~~~~~
\label{p-sol}
\ee
and substituting in (\ref{p+})--(\ref{p-bc}). This gives coupled equations for $p_0^{\pm}$ and $p_1^{\pm}$,
\be
{p_0}^\pm_{yy}=0,
\label{Order1} \\
{ p_0}^\pm_{xx}+{p_1}^\pm_{yy}=0, \label{Orderep2}
\ee
with boundary conditions 
\be
p_0^+(0,y)=1, \quad {p_0}_x^+(1,y)=0 , \quad {p_0}_y^+(x,1)=0, \label{Order1bc+} \\ p_0^-(1,y)=0, \quad {p_0}_x^-(0,y)=0, \quad {p_0}_y^- (x,-1)=0,\label{Order1bc-} 
\\
p_1^+(0,y)=0, \quad {p_1}_x^+(1,y)=0 , \quad
{p_1}_y^+ (x,1)=0, 
\label{Order_eps_sqr_bc+}
\\ 
p_1^-(1,y)=0, \quad {p_1}_x^-(0,y)=0, \quad
{p_1}_y^- (x,-1)=0. 
\label{Order_eps_sqr_bc-}
\ee
From (\ref{Order1})--(\ref{Order_eps_sqr_bc-}), we obtain $p_0^{\pm} =p_0^{\pm} (x)$ (independent of $y$ but still unknown), and
\be
{p_1}^+_{y}=(1-y){p_0}^+_{xx}, \label{p1+}\\
{p_1}^-_{y}=-(1+y){p_0}^-_{xx}. \label{p1-}
\ee
From (\ref{2flux-cty-dimless}), (\ref{vm_int}), (\ref{p1+}) and (\ref{p1-}),  we obtain the following coupled system for $p_0^\pm$,
\be 
{p_0}^+_{xx}(x)=\frac{\tilde{\Gamma} (p_0^+(x)-p_0^-(x))}{\int_{ -\frac{1}{2} }^{\frac{1}{2}  }{ \frac{d \tilde{y}}{a^4} }},
\quad p_0^+(0)=1,\quad {p_{0x}^+}(1)=0,
\label{p0+xx} \\
-{p_0}^-_{xx}(x)=\frac{\tilde{\Gamma} (p_0^+(x)-p_0^-(x))}{\int_{ -\frac{1}{2} }^{\frac{1}{2}  }{ \frac{d \tilde{y}}{a^4} }},\quad p_{0}^-(1)=0,\quad {p_{0x}^-}(0)=0,
\label{p0-xx}
\ee
where $\tilde{\Gamma}=\Gamma/(1-\delta/2)$. Note that when we further assume $\delta \ll 1 $, we may drop the tilde to give the leading order system summarized in \S\ref{modelsummary} below.

\subsubsection{Model summary}\label{modelsummary}

To summarize, we have the following asymptotic model equations, valid to leading order in $\epsilon$ and $\delta$:
\be 
{p_0}^+_{xx}(x, t)=\frac{\Gamma (p_0^+(x, t)-p_0^-(x, t))}{r_{\rm m}(x,t)},
\quad p_0^+(0, t)=1,\quad {p_0}^+_{x}(1, t)=0,
\label{p0+xx_s} \\
-{p_0}^-_{xx}(x,t)=\frac{\Gamma (p_0^+(x, t)-p_0^-(x, t))}{r_{\rm m}(x,t)},\quad p_0^-(1, t)=0,\quad {p_0}^-_{x}(0, t)=0,
\label{p0-xx_s}
\ee

\begin{equation}
\frac{\partial c (x,\tilde{y},t)}{\partial \tilde{y}}=\frac{\lambda a(x,\tilde{y},t) c(x,\tilde{y},t) r_{\rm m}(x,t)}{p_0^+(x)-p_0^-(x)} ,\quad c(x,\frac{1}{2} ,t)=1,
\label{deposition_nd_asmp_s}
\end{equation}

\begin{equation}
\frac{\partial a(x,\tilde{y},t)}{\partial t}= - c(x,\tilde{y},t), \quad a(x,\tilde{y},0)=a_0 (\tilde{y}),
\label{shrinkage_nd_s}
\end{equation}
where $r_{\rm m}(x,t)$, defined as
\be
r_{\rm m}(x, t)=\int_{ -\frac{1}{2} }^{\frac{1}{2}  }{ \frac{d \tilde{y}}{a^4(x,\tilde{y},t)} },
\label{r_m}
\ee
represents the dimensionless membrane resistance at location $x$ and time $t$. Model parameters $\Gamma$ and $\lambda$ are defined in (\ref{gamma}) and (\ref{lambda}).

\subsubsection{Method of solution}

From  (\ref{p0+xx_s}) and  (\ref{p0-xx_s}), we have ${p_0}^+_{xx}(x, t)=-{p_0}^-_{xx}(x, t)$. Integrating twice with respect to $x$ gives an expression for $p_{0}^-$ in terms of $p_{0}^+$,
\be
p_0^-(x, t)= -p_0^+ (x, t) +c_1(t)x+ c_2(t),
\label{C+-rel}
\ee
for some $c_1(t)$ and $c_2(t)$ (which are independent of $x$, but vary in time as fouling occurs). By substituting (\ref{C+-rel}) into (\ref{p0+xx}) we obtain a single equation for $p_{0}^+$ containing these two arbitrary functions of time, 
\be
r_{\rm m}(x, t){p_0}^+_{xx}(x, t)-2\Gamma{p_0^+}(x, t)=-\Gamma(c_1(t)x+c_2(t)),
\label{C-eqn}
\ee 
which must be solved subject to the four boundary conditions 
\be
p_0^+(0, t) =1, \quad p_{0x}^+ (0, t)=c_1(t), \quad p_{0}^+(1, t)=c_1(t)+c_2(t), \quad p^+_{0x}(1, t)=0.
\label{c0bc}
\ee
Hence, with $p_0^+$, $c_1$ and $c_2$ determined, we have the leading-order solution for the pressure within the support layers ($p_0^-$ can then be determined from equation (\ref{C+-rel})).

To solve the equations numerically, we discretize equations (\ref{deposition_nd_asmp_s}), (\ref{shrinkage_nd_s}) and (\ref{C-eqn}) in time and space. At each time step, we need to solve (\ref{C-eqn}) subject to the boundary conditions (\ref{c0bc}). 
The initial pore profile $a_0(\tilde y)$ is specified, so we can calculate the initial membrane resistance $r_{\rm m}$ defined in 
(\ref{r_m}), and hence solve (\ref{C-eqn}) at $t=0$ to obtain the initial (leading order) pressure distribution within the support layers. With this and the initial pore profile we can solve (\ref{deposition_nd_asmp_s}) to calculate the initial particle concentration $c$
within the membrane, and hence solve (\ref{shrinkage_nd_s}) to obtain the pore profile $a$ at the next time step.
This allows us to find the membrane resistance $r_{\rm m}$  at the new time step, and the above process repeats. We continue to solve the system until pore closure occurs at final time $t_{\rm f}$, defined to be when the flux decreases below a preset small number.\footnote{In the literature, flux falling to 10\% of the initial value is commonly used as a stopping criterion for filtration, on the practical grounds that a filter would be cleaned or discarded once the flux drops to this level, see, e.g. \cite{reis2007}. In our simulations and optimization, for simplicity we run simulations until the dimensionless flux is close to zero ($10^{-3}$).}

Note that even though pore profiles are non-uniform in $x$ during the evolution (due to the pressure gradients in $x$ that lead to differential fouling), 
pore closure ultimately occurs uniformly in $x$ along the pleat. This is because if pores at one $x$-location experience greater fouling at some time, the membrane resistance $r_{\rm m}$ at that location will be higher, so that flow is diverted to $x$-locations with lower resistance. These locations subsequently undergo increased fouling, leading to increased resistance. The net effect of this flow redistribution via resistance is that ultimately pores close uniformly in $x$ along the pleat.

\section{Results and performance optimization\label{2results}}

\subsection{Key definitions for performance evaluation \label{2optimization_definitions}}

To evaluate filter performance and to carry out design optimization, we first define some key quantities:

\be 
\mathcal{F}(t)=\int_{ 0 }^{1  }{|v_{\rm m}(x,t)|dx} \label{flux}\, ,\\
\mathcal{J}(t)=\int_{ 0 }^{t  }{{\mathcal{F}}(\tau)d\tau}\, . \label{throughput} 
\ee
 \(\mathcal{F}\)(t) is dimensionless flux, and \(\mathcal{J}\)(t) is dimensionless throughput. We note for future reference that the dimensional flux and throughput $F$ and $J$ are related to their dimensionless equivalents by the following scalings, based on (\ref{support-scalings}), (\ref{2scalings3}) and (\ref{2scalings4}): 
\be 
F(T)=\frac{K_{\rm m0}P_0 L}{\mu D} \mathcal{F}(t) \label{Flux}, \\
J(T)=\frac{K_{\rm m0}P_0 L W}{\alpha C_0 \Lambda \mu D}\mathcal{J}(t). \label{Throughput}
\ee
 We will return to these definitions when we discuss variation of the deposition parameter $\lambda$. Following earlier work (see, e.g. \cite{ho2000}), we plot  \( \mathcal{F}\) versus \( \mathcal{J}\) as one characterization of membrane performance in filtration. In our simple filtration scenario, it is desirable to keep flux high for as long as possible, while achieving a large total throughput over the filter lifetime, provided the particle removal requirement is satisfied. 
The instantaneous $x$-averaged particle concentration at the downstream side of the membrane $\tilde{y}=-1/2$, $c_{\rm avg} (t)$, another important physical quantity when considering the particle removal capability of the membrane, is defined as:
\be
c_{\rm avg}(t)=\frac{\int_{ 0 }^{1  }{|v_{\rm m}(x,t)|c(x,-\frac{1}{2},t)dx}}{ \mathcal{F}(t)}.
\ee
This quantity represents the averaged particle concentration in the filtrate\footnote{Note that ``filtrate'' here refers to the feed that has passed through the filter.}, sampled at any given time.  We also monitor the accumulated particle concentration of the filtrate, $c_{\rm acm} (t)$, defined as:
\be
c_{\rm acm}(t)=\frac{\int_{0}^{t}\int_{ 0 }^{1  }{|v_{\rm m}(x,\tau)|c(x,-\frac{1}{2},\tau)dxd\tau}}{\mathcal{J}(t)}.
\ee 
This quantity represents the particle concentration of the accumulated filtrate at any given time (suppose all the filtrate is collected in a well-mixed jar as the filtration progresses, then $c_{\rm acm} (t)$ is the particle concentration of the collected filtrate in the jar at time $t$).

\subsection{Formulating the optimization problem\label{2optimization_formulation}}

Although we can predict the performance for any given initial pore profile distribution using our model (\ref{p0+xx_s})--(\ref{shrinkage_nd_s}), a question of greater interest is to find the {\it optimized} pore profile for any given filtration objective and operating conditions  (referred to as the objective or cost function and constraints, respectively, in the optimization literature). 
For our case, we consider the optimization process as finding the pore profile that gives  
{\it  the highest total throughput } \( \mathcal{J}\)$(t_{\rm f})$ ((\ref{throughput}) evaluated at $t_{\rm f}$) while maintaining the {\it initial particle removal threshold $R$ above a certain percentage} at time $t=0$ ($c_{\rm avg}(0) \leq 1-R/100$). In our example simulations that follow, the particle removal threshold is set at $R=99\%$, or $R=99.9\%$. 

The general optimization problem is challenging (it is in general not convex, and the searching space is infinite dimensional), even for the simple model presented here. We here present results for optimizing only within the limited class of initial pore profiles $a_0(\tilde{y})$ represented in terms of (low-degree) polynomials. We vary the coefficients of polynomials to find the values that maximize \( \mathcal{J}\)$(t_{\rm f})$ while satisfying the particle removal threshold constraint. We used the  {\tt MultiStart}
algorithm with {\tt fmincon}
  as local solver from the {\tt  MATLAB}\textregistered~{\tt Global Optimization} toolbox for this optimization. {\tt MultiStart} uses uniformly-distributed or user-supplied start points within the searching domain to perform repeatedly gradient descent to find local minimizers of the cost function. The user has control over the number of searching points (start points) the {\tt MultiStart} algorithm uses for running the local solver. After {\tt MultiStart} reaches the stopping criterion, the algorithm creates a vector of {\tt GlobalOptimSolution} objects,  which contains the global minimum of the objective (cost) function and the minimizer that gives this global minimum (here, the optimal coefficients of the polynomial describing the pore profile). In our case, the cost function is defined as the negative of  \( \mathcal{J}\)$(t_{\rm f})$ (found by evaluating (\ref{throughput}) at the pore closure time $t_{\rm f}$) provided neither the particle removal threshold constraint nor the physical constraint (pore size cannot exceed the dimensions of the containing box or be negative, i.e. $0 < a \le 1$) is violated; the cost function takes value 1 if the physical constraint is violated and it takes value 2 if the particle removal threshold constraint is violated:\footnote{The physical constraint is checked first. The threshold constraint will not be checked if the physical constraint is violated already.} 
\be
\text{Cost function}= \left\{
\begin{array}{ll}
-\text{ \( \mathcal{J}\)$(t_{\rm f})$} & \text{if no constraint is violated,}\\
1  & \mbox{if physical constraint is violated,}\\
2  & \text{if threshold constraint is violated.}
\end{array}
\right.
\label{2objectivefun}
\ee

\subsection{Simulation results for the optimization \label{results}}

Since fouling is an integral part of the filtration process, any factors that influence fouling will automatically impact the optimization. Three principal factors are identified in the literature as affecting fouling \citep{tang2011}: (i)
{\it feed characteristics}, (ii) {\it membrane properties} and (iii) {\it operational conditions}. These factors are all represented in our model parameter $\lambda$ (\ref{lambda}), which depends on $\mu$ (feed viscosity; {\it feed characteristics}); $\Lambda$ (the particle/pore attraction coefficient, a function of both {\it feed characteristics} and {\it membrane properties}); $D$, $W$, $K_{\rm m0}$ (membrane thickness, size of pore-containing box, membrane permeability; {\it membrane properties}); and $P_0$ (applied pressure drop; {\it operational conditions}). Larger values of $\lambda$ indicate a system with superior particle capture efficiency: particles are more easily captured by the membrane compared to systems with smaller $\lambda$ values. We present results for two values: $\lambda=0.1$ and $\lambda=1$. The model parameter $\Gamma$ (\ref{gamma}) also characterizes certain membrane properties, but throughout our simulations we hold its value fixed (Table~\ref{2t:parameters2}).

Depending on which physical quantity induces the change in $\lambda=(\pi \mu D^2 \Lambda)/(4 W K_{\rm m0} P_0)$, a scaling factor for the dimensionless flux and throughput,  \( \mathcal{F}\) and  \( \mathcal{J}\), may be necessary to make direct performance comparisons between simulations with different $\lambda$-values, see (\ref{Flux}), (\ref{Throughput}). For example, if a factor of 10 increase in the dimensional deposition coefficient $\Lambda$ induces the change from $\lambda=0.1$ to $\lambda=1$, then the dimensionless flux for $\lambda=1$ must be decreased by a factor of 10 when comparing with the dimensionless flux for $\lambda=0.1$ to give a proper comparison. However, if $\lambda$ is changed via increasing membrane thickness $D$ and pressure $P_0$ by the same factor, while keeping ${K_{\rm m0}P_0 L}/{(\mu D)}$ and ${K_{\rm m0}P_0 L W}/{(\mu D \alpha C_0 \Lambda)}$ constant, then the dimensionless plots as presented here will be a proper comparison.

\begin{figure}
\hspace{0.3cm}{\scriptsize (a)}\rotatebox{0}{\includegraphics[width=0.44\linewidth, height=0.39\linewidth]{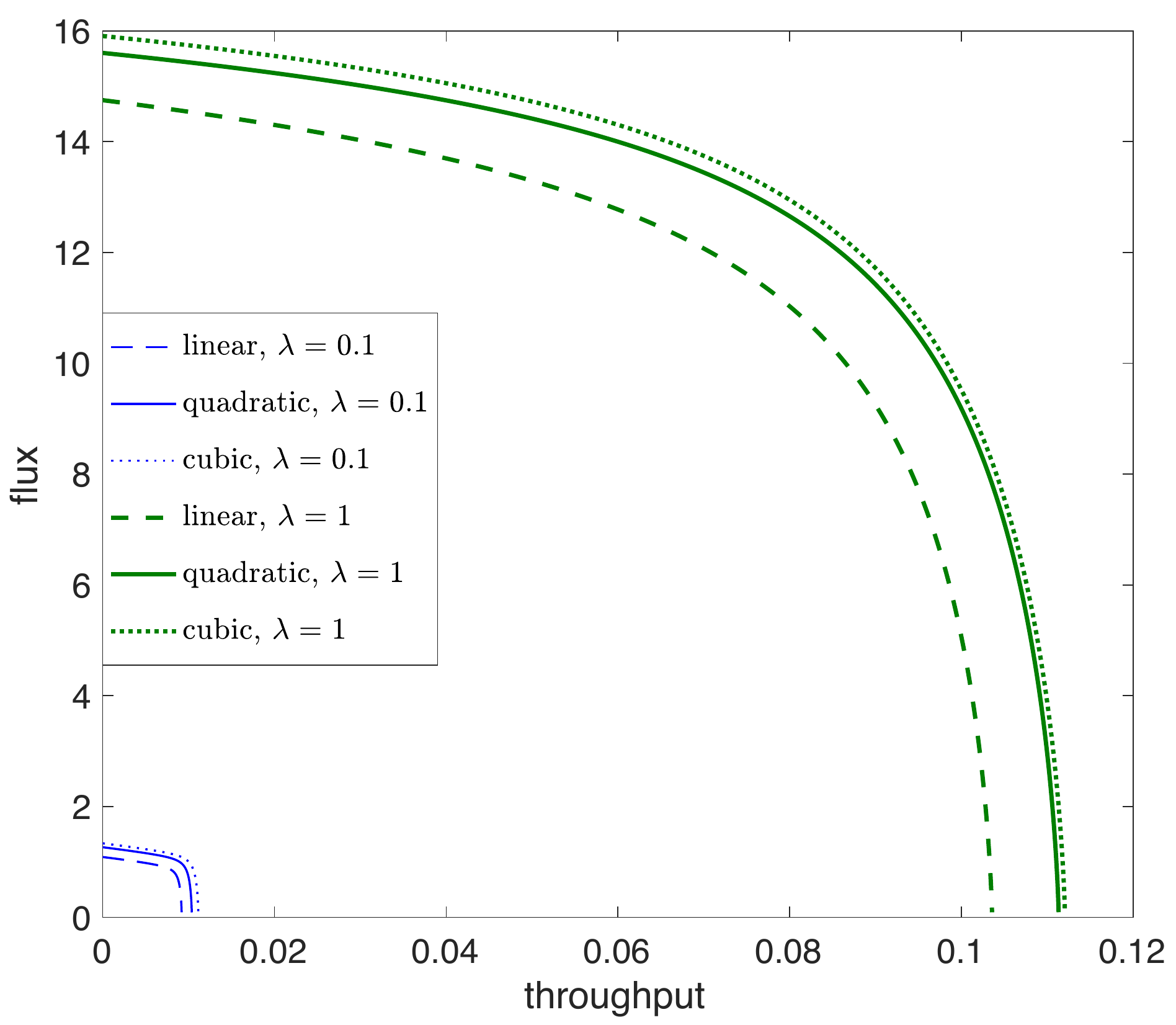}}\hspace{0.2cm}{\scriptsize (b)}\includegraphics[width=0.445\linewidth, height=0.39\linewidth]{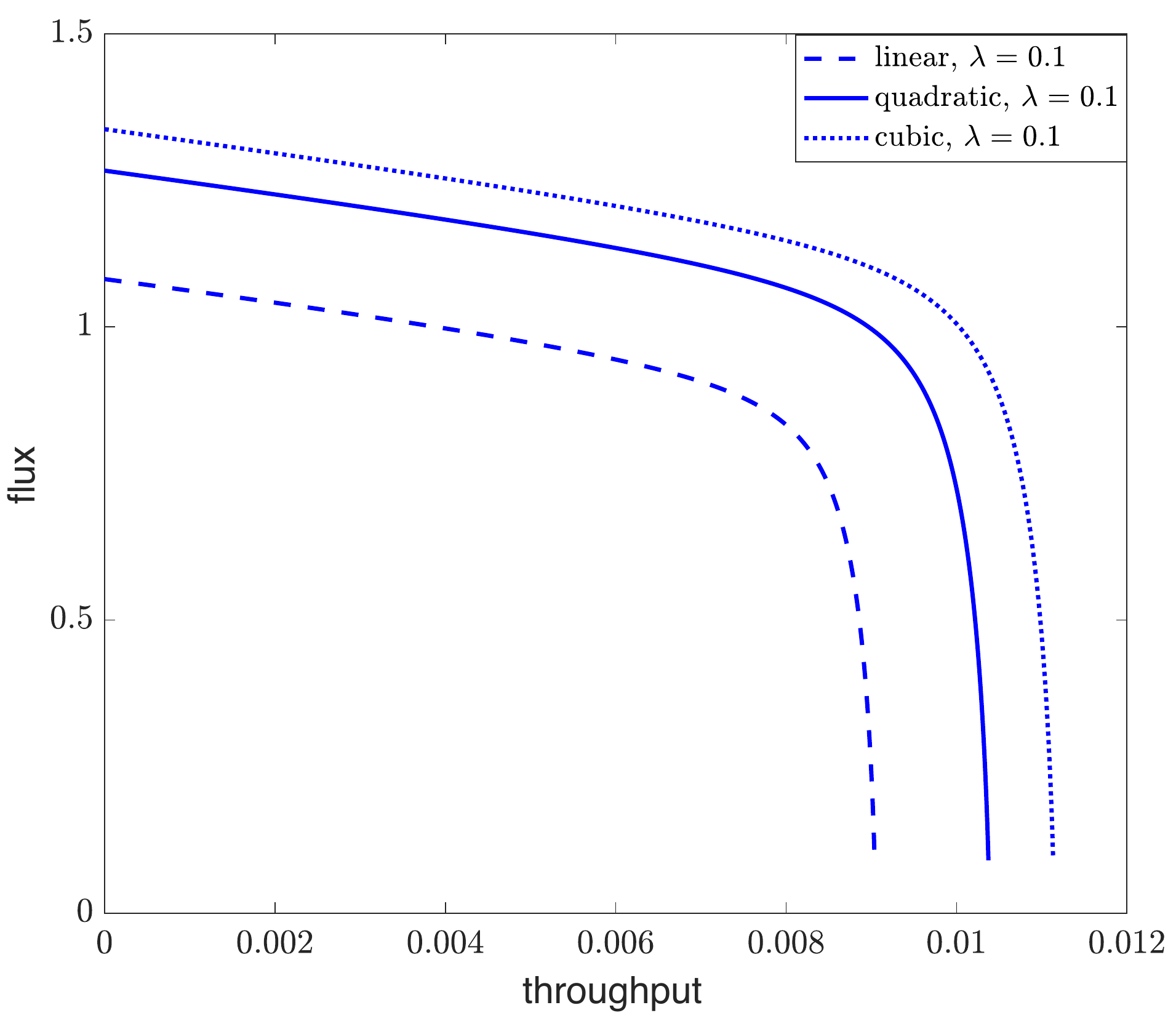} 
{\scriptsize (c)}\includegraphics[width=0.465\linewidth, height=0.39\linewidth]{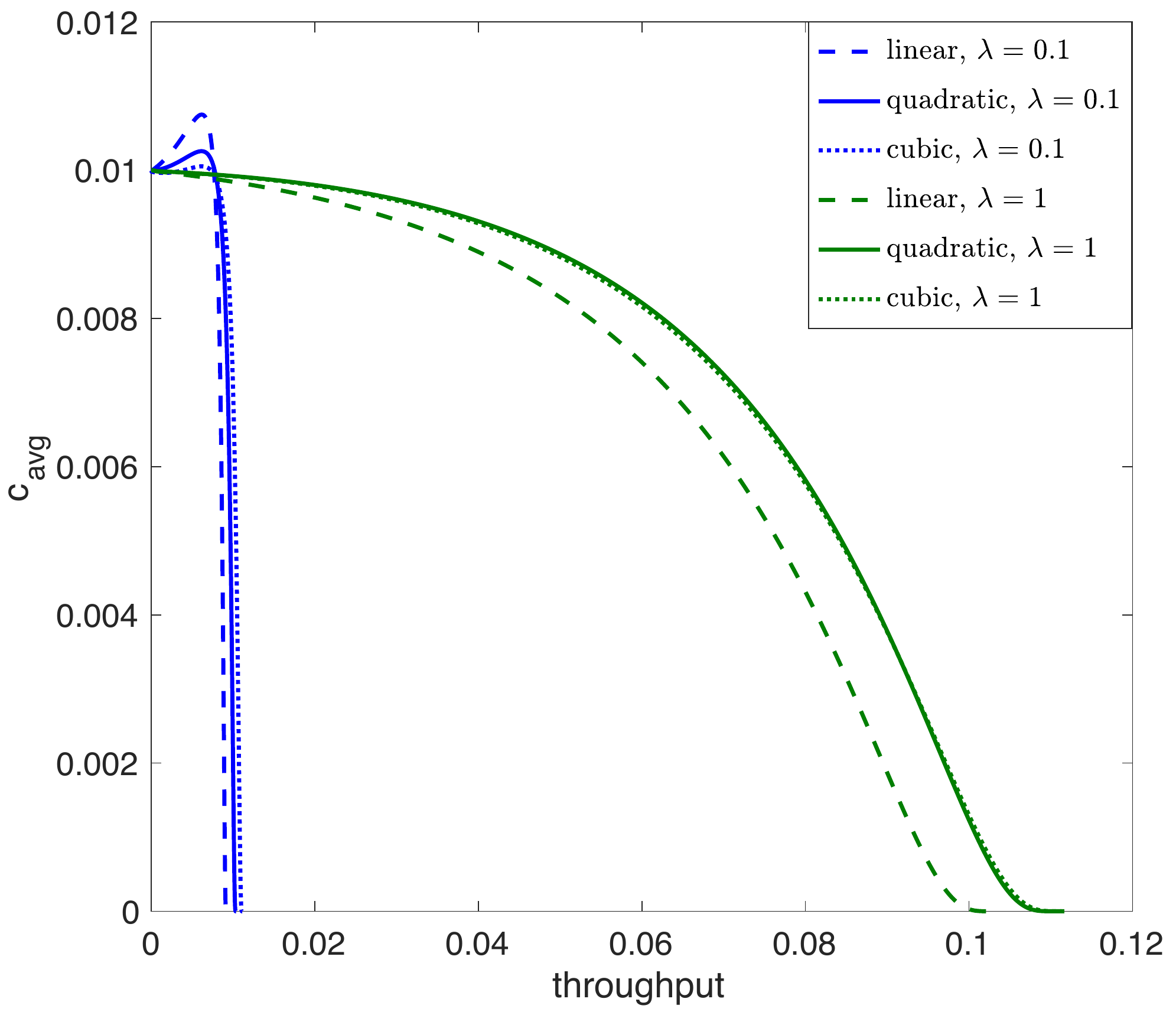}{\scriptsize (d)}\includegraphics[width=0.465\linewidth, height=0.39\linewidth]{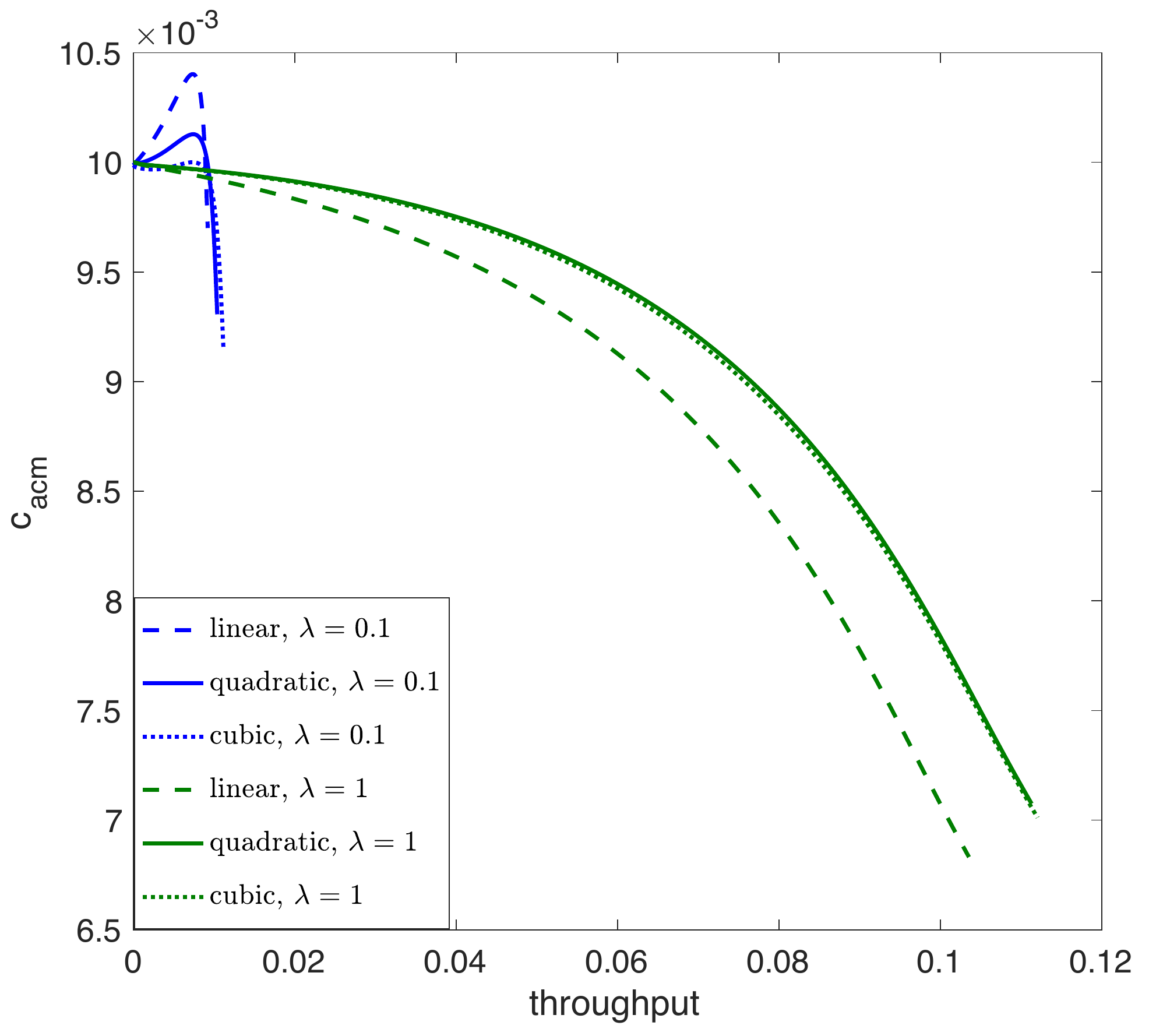}
{\scriptsize (e)}\includegraphics[width=0.465\linewidth, height=0.39\linewidth]{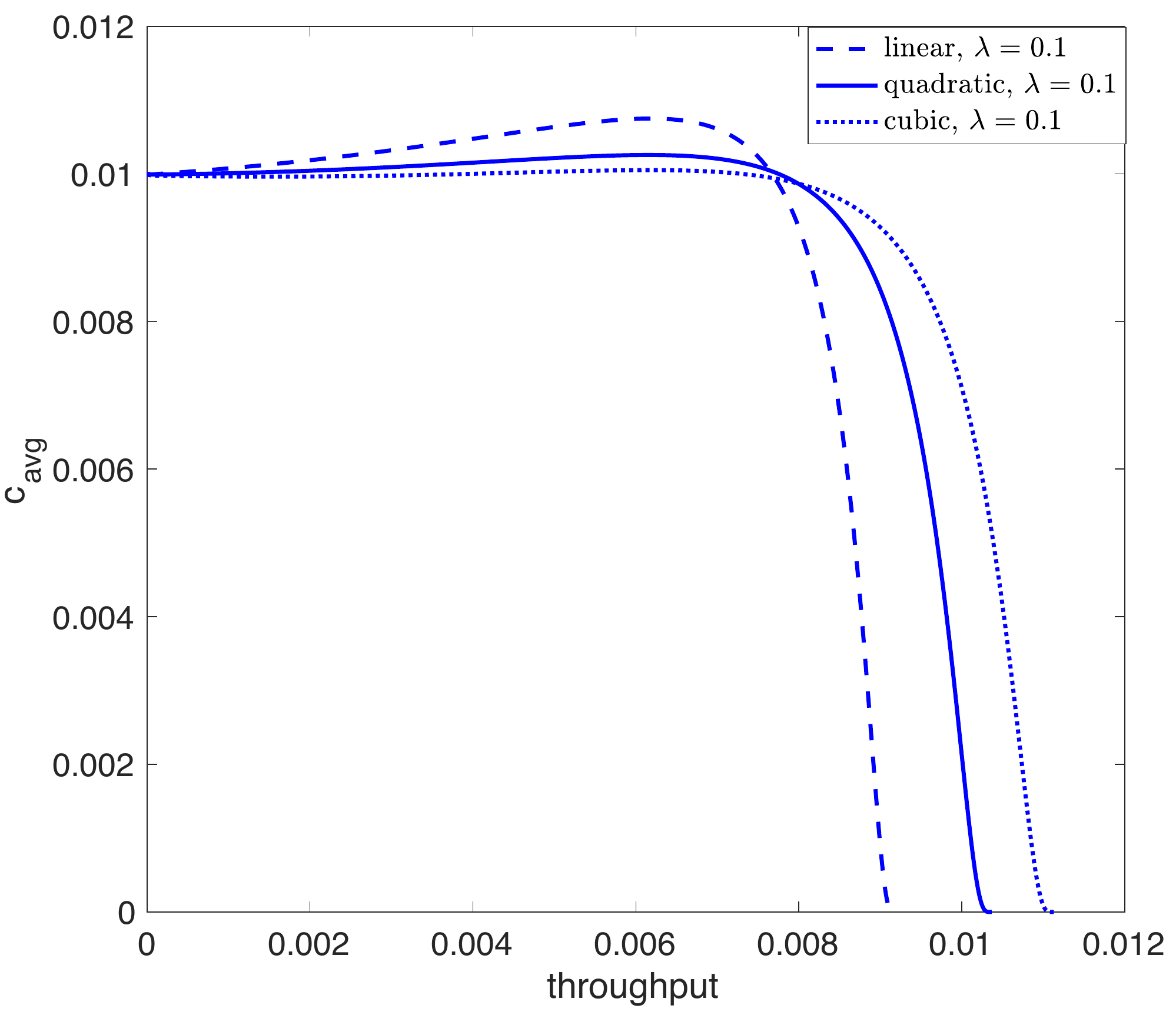}\hspace{0.1cm}{\scriptsize (f)}\includegraphics[width=0.465\linewidth, height=0.39\linewidth]{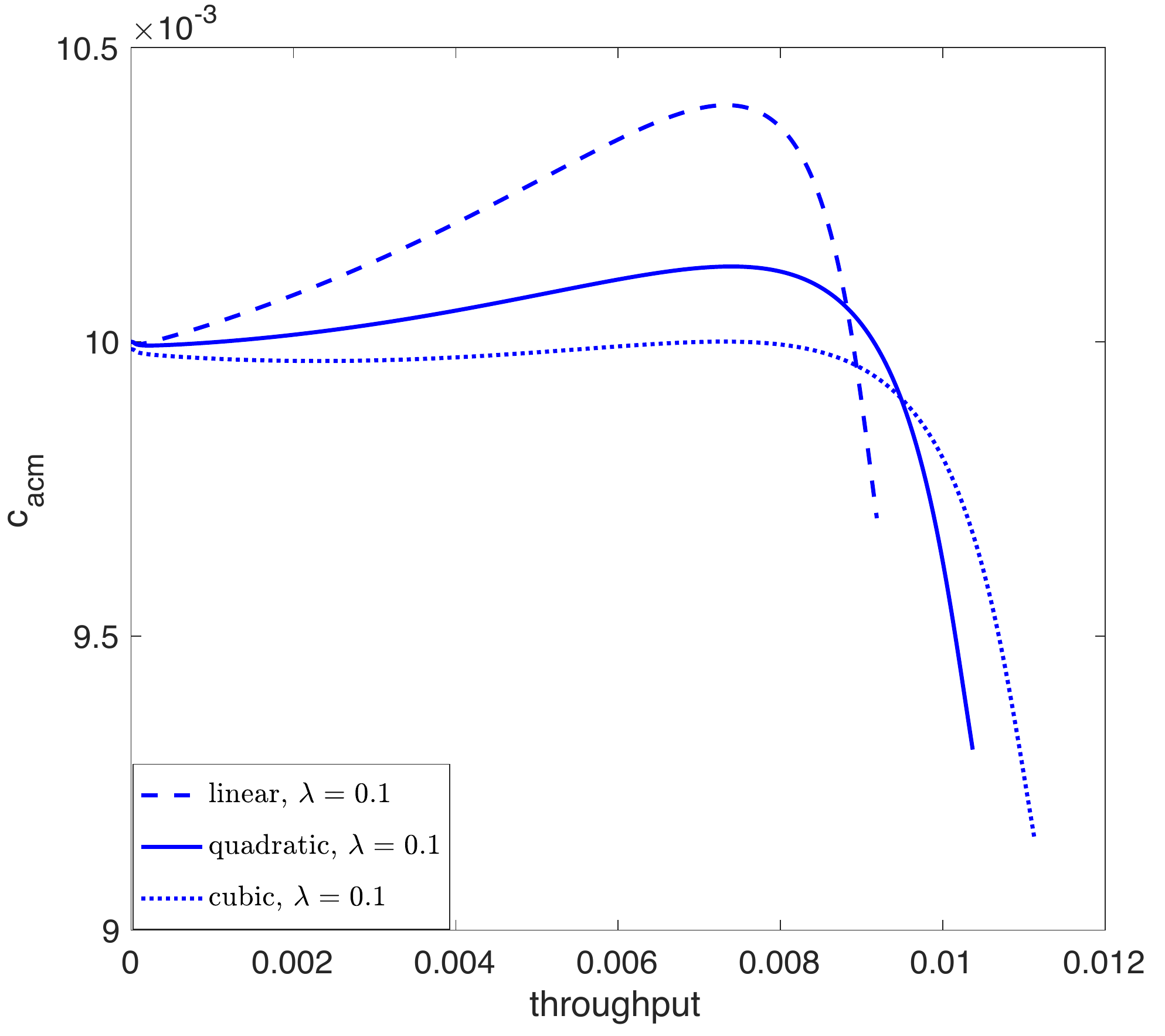}
\caption{\footnotesize{Results with particle removal threshold $R$ fixed at 99\%, $\lambda=0.1$ in blue, and $\lambda=1$ in green, for optimal pore profiles from the classes of linear, quadratic and cubic polynomials: (a) Flux-throughput (\( \mathcal{F}\)- \( \mathcal{J}\)) plots for all cases;
  (b) Zoomed flux-throughput (\( \mathcal{F}\)- \( \mathcal{J}\)) plot for $\lambda=0.1$; (c) Instantaneous averaged particle concentration  $c_{\rm avg}$ vs. throughput  \( \mathcal{J}\);
(d) Accumulated average particle concentration  $c_{\rm acm}$ vs. throughput  \( \mathcal{J}\); (e) Zoom of the $\lambda=0.1$ results from (b); (f) Zoom of the $\lambda=0.1$ results from (d). 
}
} 
\label{2fig_optimization_a_0_y}
\end{figure}

Figure \ref{2fig_optimization_a_0_y} shows results for the optimization procedure outlined in \S\ref{2optimization_formulation} above. 
Optimization is carried out for initial pore profiles $a_0(\tilde{y})$ within the classes of linear, quadratic and cubic polynomials in $\tilde{y}$ (distinguished by the linestyles), with the particle removal threshold constraint fixed at $R=99\%$. As one would expect, increasing the order of the polynomial that describes the initial pore profile leads to increased throughput \( \mathcal{J}\), since we are optimizing over a larger function class. However, gains are minimal between the quadratic and cubic cases, indicating (a) rapid convergence towards the global optimum, and (b) that optimizing over low-order polynomials ({\it e.g.} quadratics) may be sufficient for practical purposes.

Results for $\lambda=0.1$ (blue curves), and $\lambda=1$ (green curves) are shown, for a total of six optimized scenarios. Figure \ref{2fig_optimization_a_0_y}(a) shows the  \( \mathcal{F}\)-\( \mathcal{J}\) curves for the membrane with the optimized pore profiles in all six cases.
Figure \ref{2fig_optimization_a_0_y}(b) shows a magnified version of these plots for $\lambda=0.1$, which are difficult to distinguish in (a). We observe that for fixed particle removal threshold ($R=99\%$), increasing the dimensionless particle retention coefficient $\lambda$ gives higher total dimensionless throughput. To put these results in context, consider the two specific scenarios outlined above. In the first, if $\lambda$ is changed by changing $\Lambda$, then in dimensional terms the two scenarios would give rather comparable outcomes in terms of \( \mathcal{J}\)$(t_{\rm f})$. However, the flux for the larger $\lambda$-value is always significantly higher, indicating that the filtration process would be of much shorter duration. This can be explained by the fact that a stickier membrane can remove the same fraction of particles with much larger pores, admitting a larger flux. The larger flux will lead to faster fouling, leading to a shorter filtration time to obtain the same total throughput (which in most applications would be considered desirable). In the second scenario $\lambda$ is changed by changing both membrane thickness $D$ and driving pressure $P_0$ by the same factor. Here the dimensionless  \( \mathcal{F}\)- \( \mathcal{J}\) plots may be directly compared, and we see that the larger $\lambda$-value is clearly superior. Two competing factors are at play here: on the one hand driving pressure is higher, which favors higher flux; but on the other hand the increased membrane thickness imparts higher system resistance, tending to lower the flux. The increased membrane depth allows the particle removal requirement to be satisfied with considerably larger pores, however, so that the net effect of the increased pressure outweighs that of the thicker membrane, leading to much higher flux and throughput over the filtration duration.


Figures \ref{2fig_optimization_a_0_y}(c) and (d) show the instantaneous and accumulated particle concentrations in the filtrate,  $c_{\rm avg}$ and $c_{\rm acm}$ respectively, versus throughput  \( \mathcal{J}\), for the corresponding optimized membranes (note that  \( \mathcal{J}\) is a monotonic increasing function of time $t$ with $\mbox{ \( \mathcal{J}\)}(0)=0$, so the same trends will be observed as when $c_{\rm avg}$ is plotted against time $t$). These figures confirm that the particle removal constraint is met precisely at the start of the filtration, as one would expect. However, 
in figure \ref{2fig_optimization_a_0_y}(c) we observe that the averaged particle concentration for $\lambda=0.1$ (blue curves) {\it increases} for a certain period of time before decreasing again. This concentration increase was also observed in the corresponding accumulated particle concentration plot in figure \ref{2fig_optimization_a_0_y}(d).  This behavior is undesirable for particle removal applications: users would like to be assured that particle retention by the membrane will not deteriorate during the course of filtration.  To the best of our knowledge this phenomenon was not observed in earlier models of adsorptive fouling; however, it has been reported experimentally ({\it e.g.} \cite{jackson2014,lee2019}) and in models of fouling by total blocking~\citep{beuscher2010}. 
This issue will be further discussed in the Conclusions. Observe also that the case $\lambda=1$ leads to a much more dilute concentration of impurities in the filtrate, evidenced by the lower value of $c_{\rm acm}$ at the conclusion of the filtration (figure~\ref{2fig_optimization_a_0_y}(d)). Figures \ref{2fig_optimization_a_0_y}(e) and (f) show the zoom-in of the $\lambda=0.1$ results from figures \ref{2fig_optimization_a_0_y}(c) and (d) respectively, for better visibility. 

\begin{figure}
{\scriptsize (a)}\rotatebox{0}{\includegraphics[width=0.465\linewidth, height=0.40\linewidth]{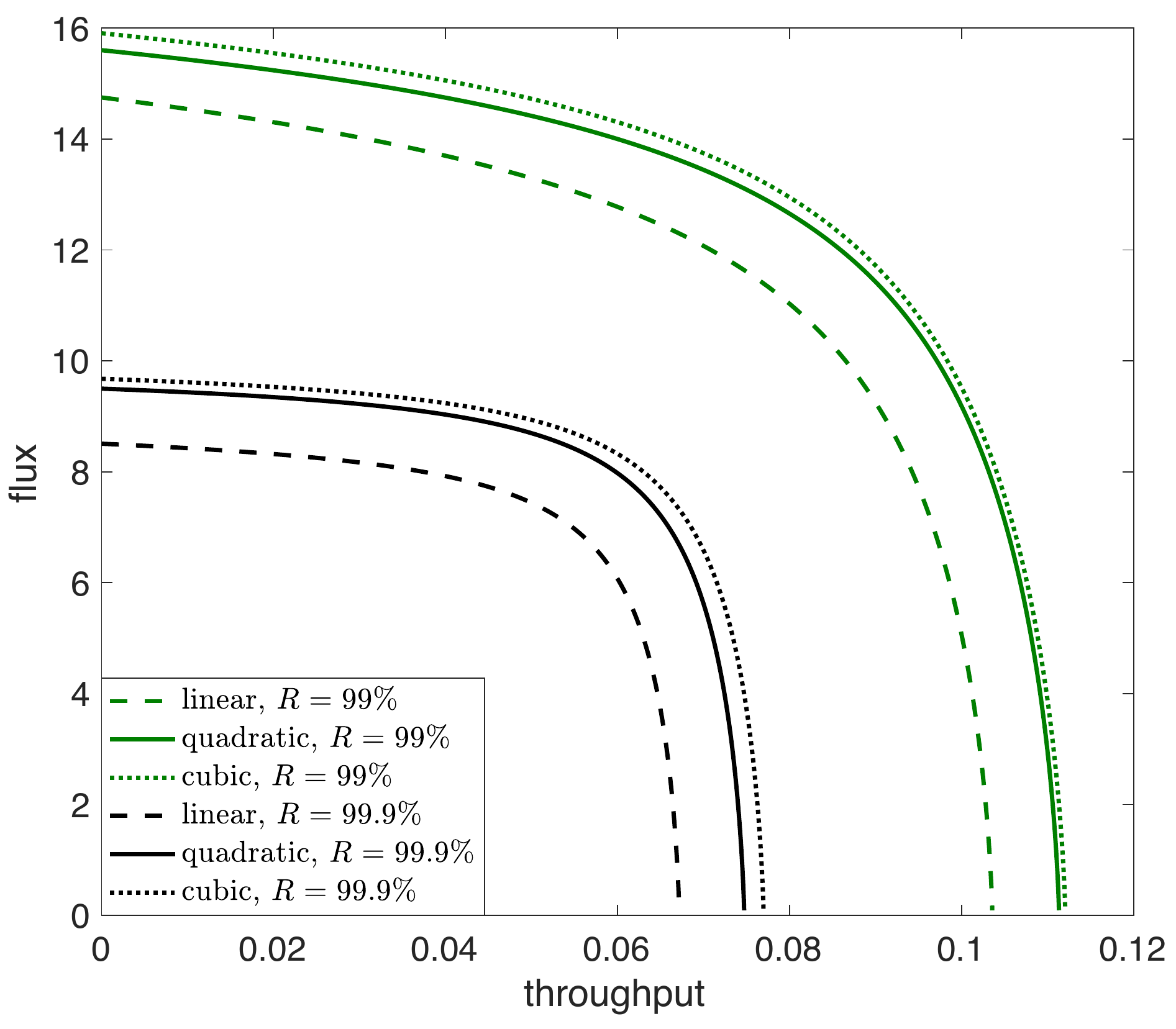}}{\scriptsize (b)}\includegraphics[width=0.465\linewidth, height=0.40\linewidth]{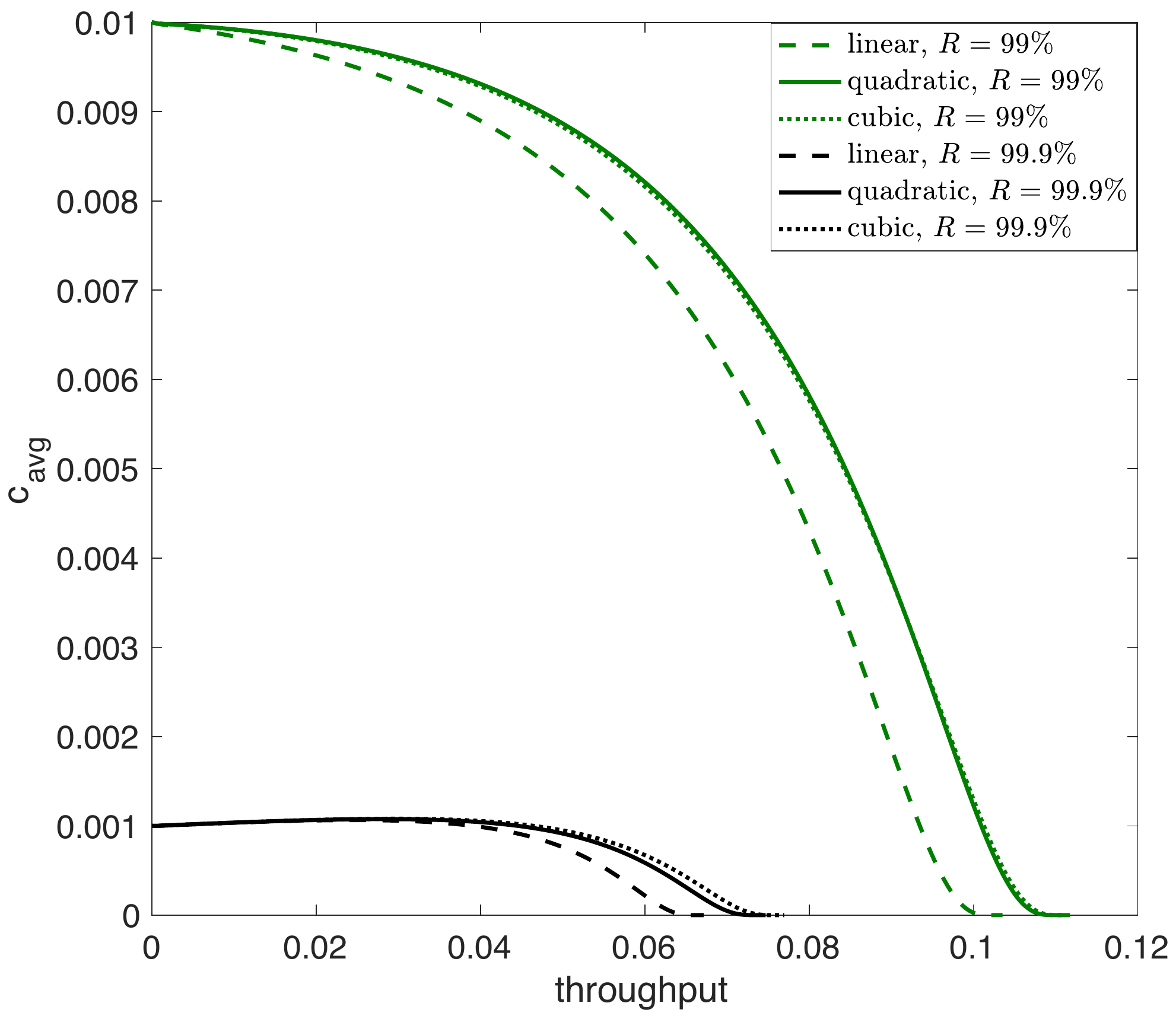} 
{\scriptsize (c)}\includegraphics[width=0.465\linewidth, height=0.40\linewidth]{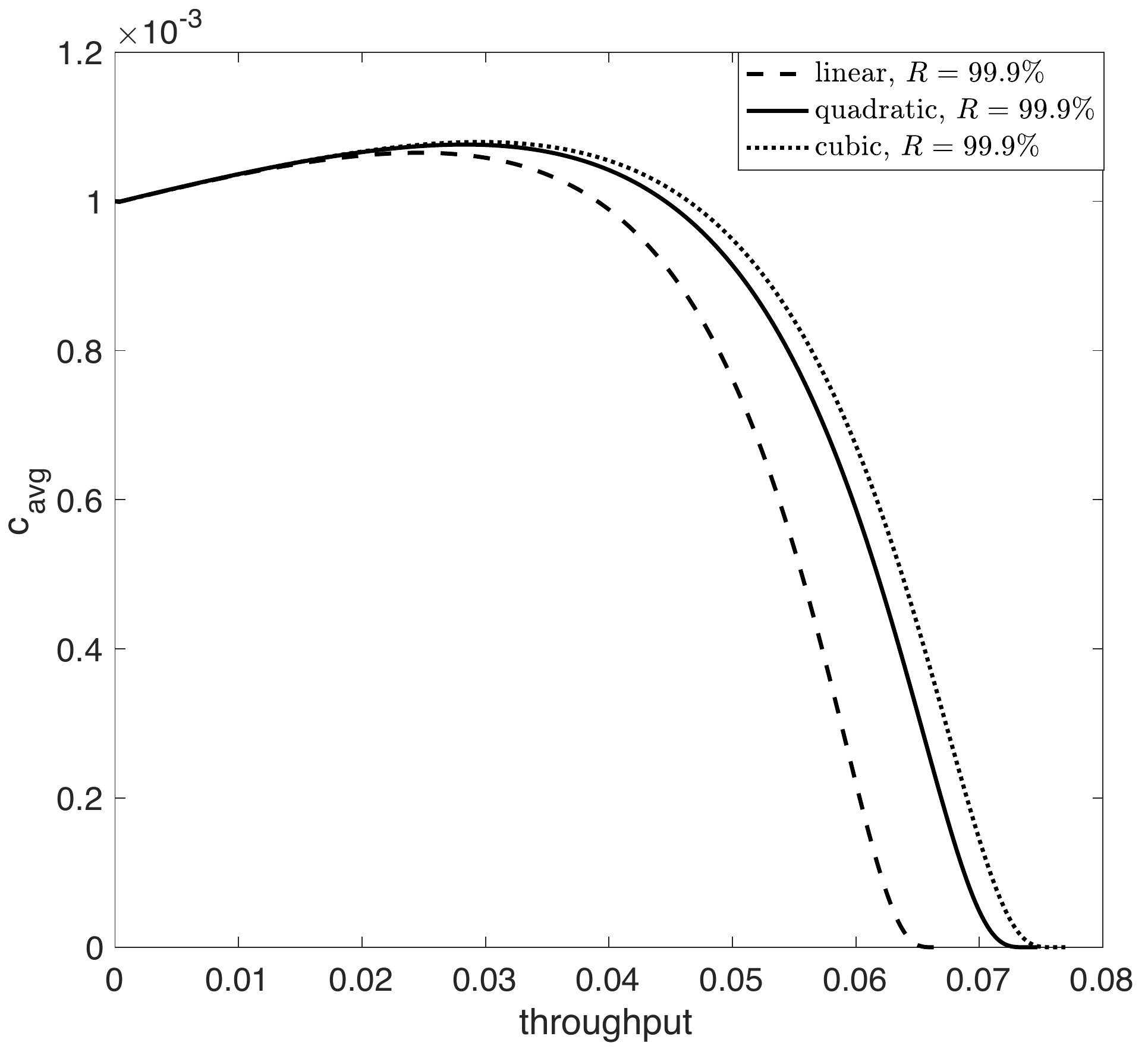}{\scriptsize (d)}\includegraphics[width=0.465\linewidth, height=0.40\linewidth]{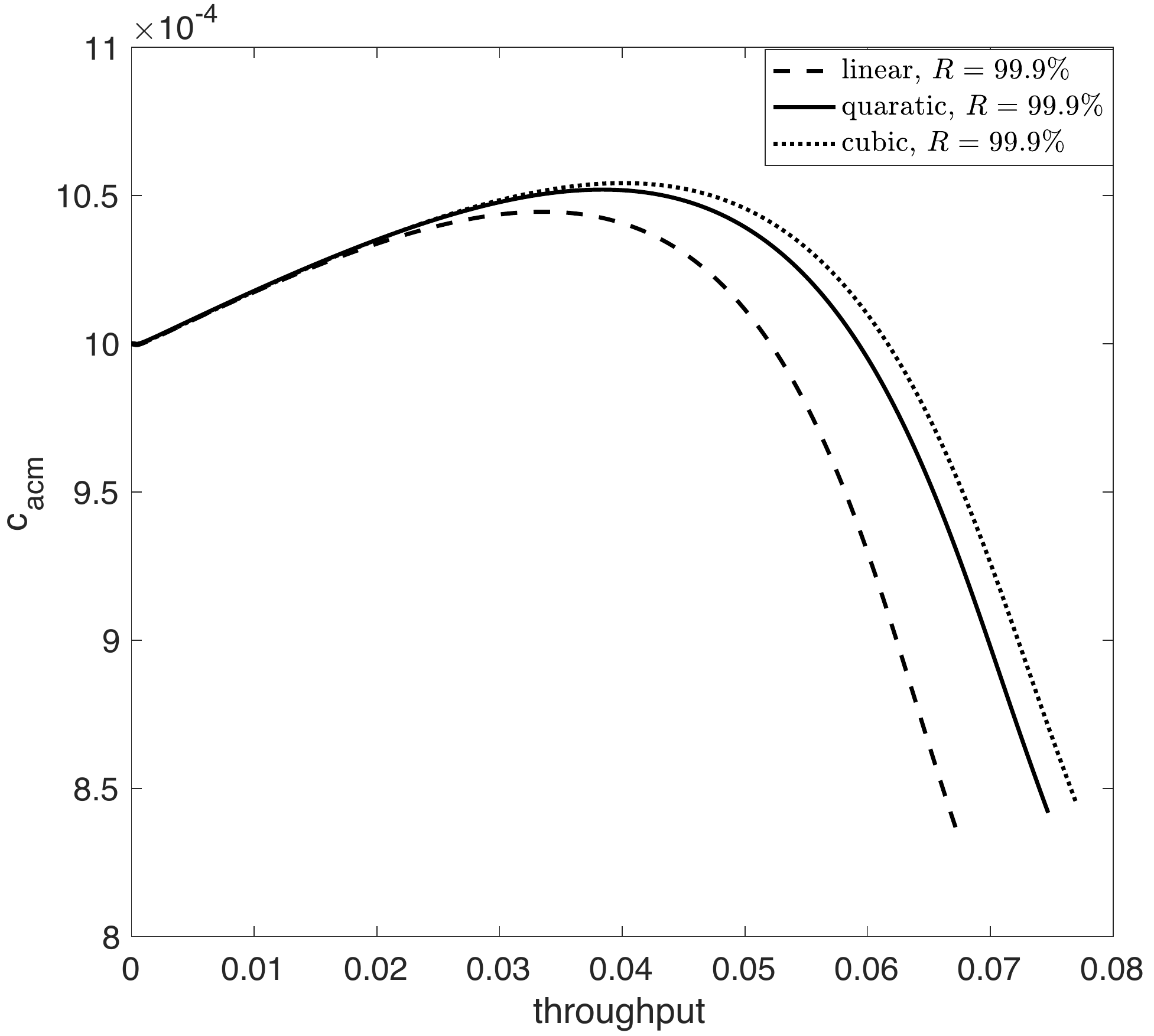}
\caption{\footnotesize{Results for membranes with optimized pore profiles from the classes of linear, quadratic and cubic polynomials, with $\lambda=1$ and $R=99\%$ (green curves) or $R=99.9\%$ (black curves): (a) flux-throughput (\( \mathcal{F}\)- \( \mathcal{J}\)) plot for all cases; (b) Instantaneous averaged particle concentration  $c_{\rm avg}$ vs. throughput  \( \mathcal{J}\) in the filtrate for the optimized profiles; (c) Zoom of the $R=99.9\%$ results from (b); (d) Accumulated average particle concentration  $c_{\rm acm}$ vs. throughput \( \mathcal{J}\) in the filtrate for the optimized profiles with $R=99.9\%$.}} 
\label{2fig_optimization_a_0_y_var_R}
\end{figure}

Different applications may require different particle removal thresholds $R$. In figure \ref{2fig_optimization_a_0_y_var_R} we explore how changing $R$ affects the optimized results, with $\lambda$ fixed at $\lambda=1$. Figure \ref{2fig_optimization_a_0_y_var_R}(a) shows the  \( \mathcal{F}\)- \( \mathcal{J}\) plots for three optimized membranes (again with pore profiles optimized over the classes of linear, quadratic, and cubic polynomials), for $R=99\%$ (green; these curves are the same as the green curves in figure~\ref{2fig_optimization_a_0_y}(a)) and $R=99.9\%$ (black).  It is observed that if we insist on removing more particles ($R=99.9\%$), we reduce the instantaneous flux and total throughput. This makes sense since the stricter requirement for particle removal (while keeping $\lambda$ fixed) necessitates that pores be narrower to reduce the number of particles that can pass through the membrane, which decreases both \( \mathcal{F}\) and  \( \mathcal{J}\)$(t_{\rm f})$. Figure \ref{2fig_optimization_a_0_y_var_R}(b) shows the averaged instantaneous particle concentration $c_{\rm avg}$ versus throughput \( \mathcal{J}\) in the filtrate for the corresponding optimized profiles, showing that again the particle removal thresholds are met precisely at $t=0$ (the green curves for $R=99\%$ are the same as those in figure~\ref{2fig_optimization_a_0_y}(c)). Figure \ref{2fig_optimization_a_0_y_var_R}(c) shows the zoom-in of these plots for $R=99.9\%$, since these curves are hard to distinguish in \ref{2fig_optimization_a_0_y_var_R}(b). This zoomed plot reveals that once again $c_{\rm avg}$ may increase over the course of the filtration, suggesting that a high particle retention requirement may lead to the risk of decreased particle retention capability later on. Figure \ref{2fig_optimization_a_0_y_var_R}(d) shows the accumulated particle concentration $c_{\rm{acm}}$ in the filtrate for the optimized profiles with the higher removal threshold $R=99.9\%$. Again, the deterioration in particle retention over time is apparent.

\begin{figure}
\centering
{\scriptsize (a)}\rotatebox{0}{\includegraphics[scale=.315]{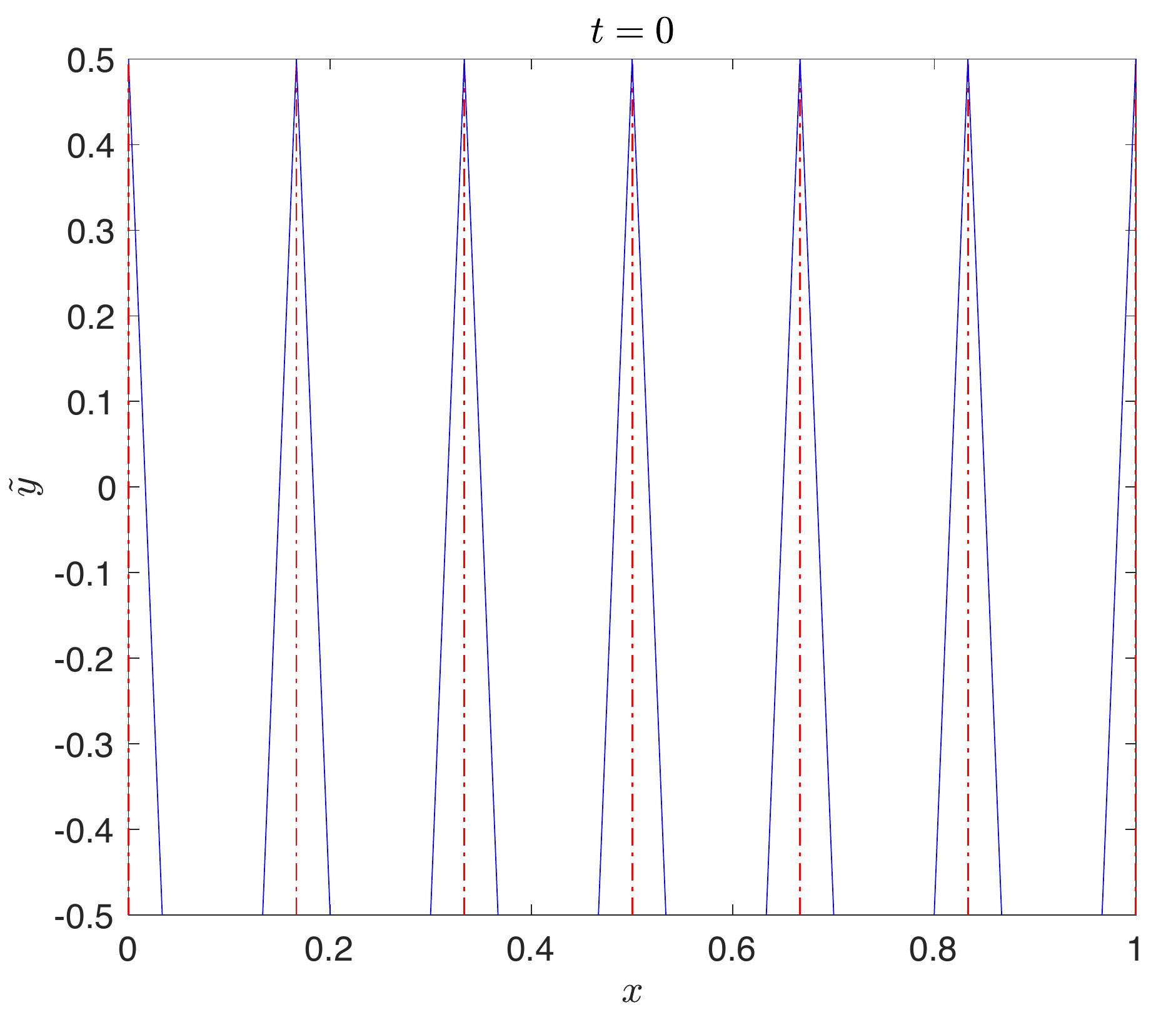}}
{\scriptsize (b)}\includegraphics[scale=0.32]{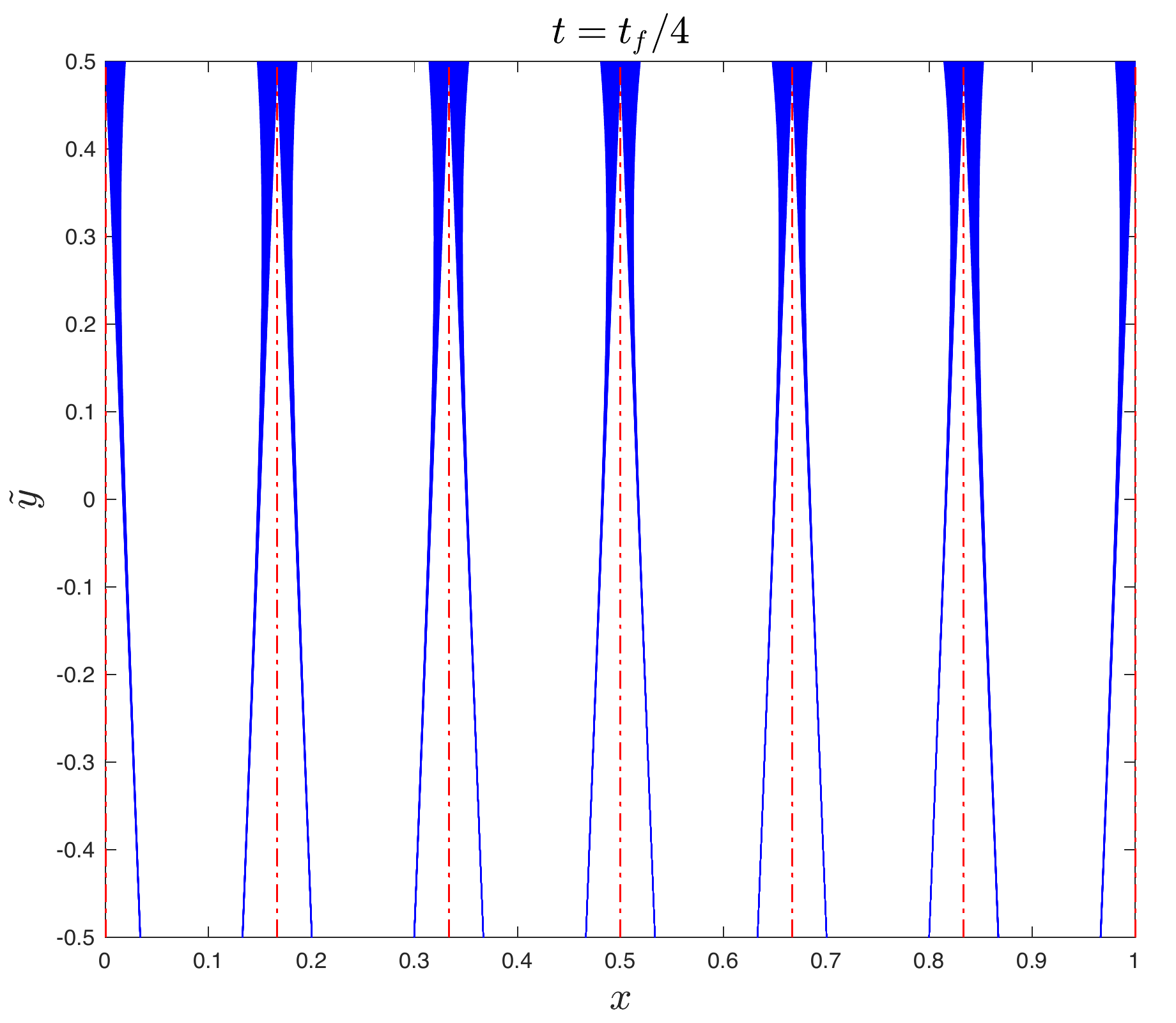}
{\scriptsize (c)}\includegraphics[scale=0.315]{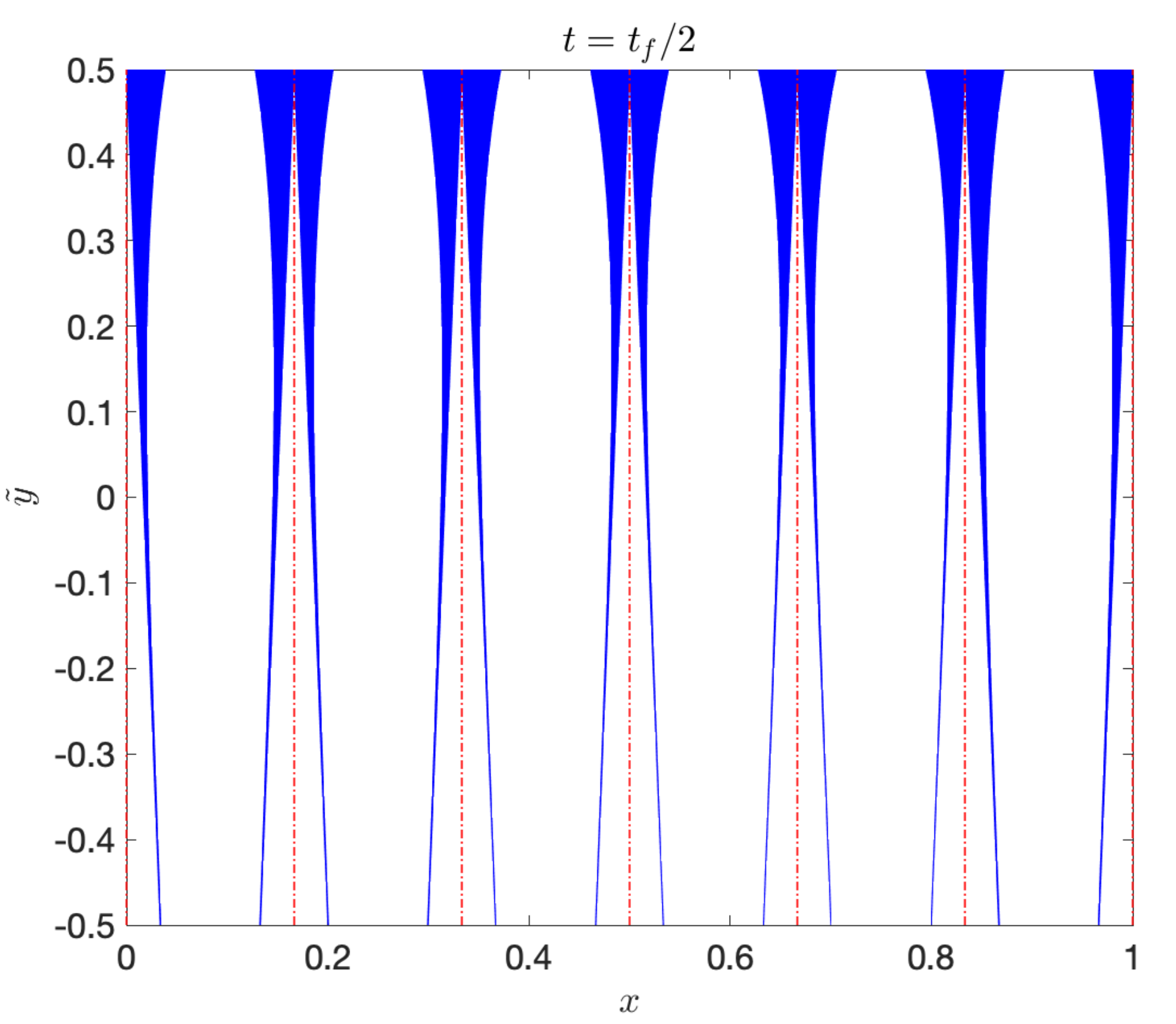}
{\scriptsize (d)}\includegraphics[scale=0.32]{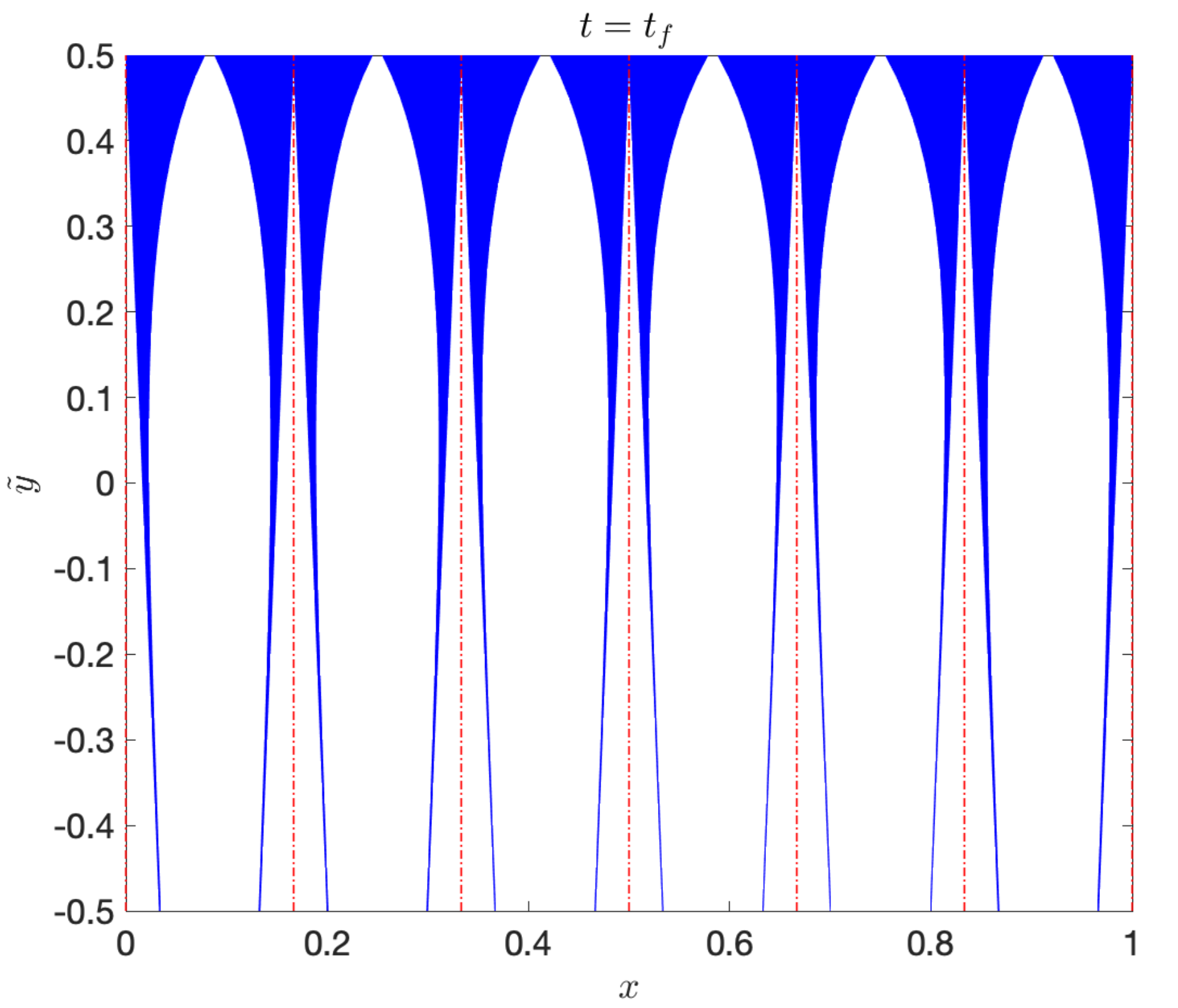}
\vspace{0.5cm}

{\scriptsize (e)}\rotatebox{0}{\includegraphics[scale=1.5]{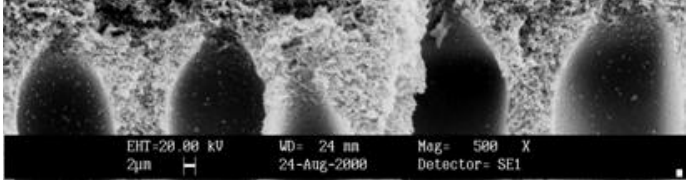}}
\caption{\footnotesize{ (a)--(d): Optimized linear pore profile evolution at different $x$-locations, with $\lambda=1, R=99\%$, red dashed line indicating the boundary of the prism containing each pore, and the solid blue color indicating the cumulative particle deposition on the pore wall: (a) $t=0$, (b)  $t=t_{\rm f}/4$, (c)  $t=t_{\rm f}/2$, (d)  $t=t_{\rm f}$; (e) experimental result showing heavily-fouled ultrafiltration membrane in water treatment system. The black pore regions are clearly visible \citep{sun2019}.}}
\label{2fig_optimization_a_0_y_evo}
\end{figure}

For completeness, figure \ref{2fig_optimization_a_0_y_evo} illustrates typical pore evolution during the filtration at different $x$-locations along the pleat, using the optimized linear pore profile obtained with $\lambda=1$, $R=99\%$. The evolution of six selected pores is shown, in cross-section (recall that the pores as modeled are three dimensional; see the schematic in figure \ref{2idealized}(b)). The red dashed line indicates the boundary of the pore-containing box. The blue lines in (a) indicate the optimal (linear) pore walls, and in (b,c,d) the increasing blue area indicates the particles that have deposited (fouling) on the pore walls shrinking the pore radius. As explained earlier, pore closure is observed to be a self-leveling process: at the final time, all pores close simultaneously. Figure \ref{2fig_optimization_a_0_y_evo}(e) shows an experimental image from a real track-etched membrane, showing good qualitative agreement with our simulations. In addition, we note that (though it may not be obvious from figure \ref{2fig_optimization_a_0_y_evo}) for simulations such as these, where there is no $x$-variation in the initial pore profiles, the evolving pore profiles retain symmetry about $x=1/2$ (the point mid-way between pleat tips and valleys), with $a(x,\tilde{y},t)=a(1-x,\tilde{y},t)$ for $0\leq x\leq 1$, $-1/2\leq \tilde{y}\leq 1/2$, $0\leq t\leq t_{\rm f}$. In Appendix \ref{symmetry}, we demonstrate that this is true more generally, with initial pore profile distributions symmetric about $x=1/2$ retaining this symmetry throughout the evolution.
   
\subsection{Simulation results: Pore size variability\label{variability}}

\begin{figure}
{\scriptsize (a)}\rotatebox{0}{\includegraphics[scale=.33]{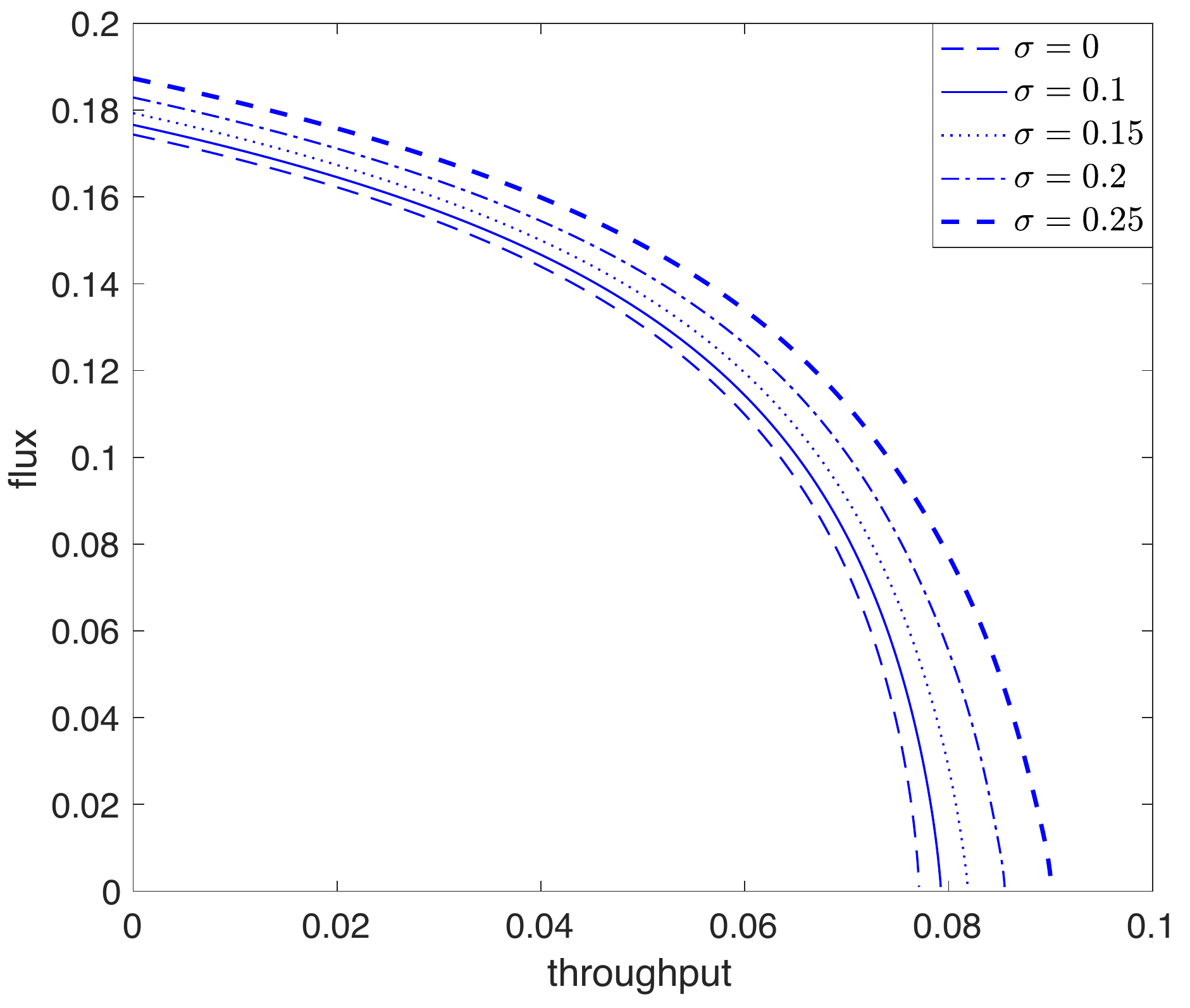}}
{\scriptsize (d)}\includegraphics[scale=0.335]{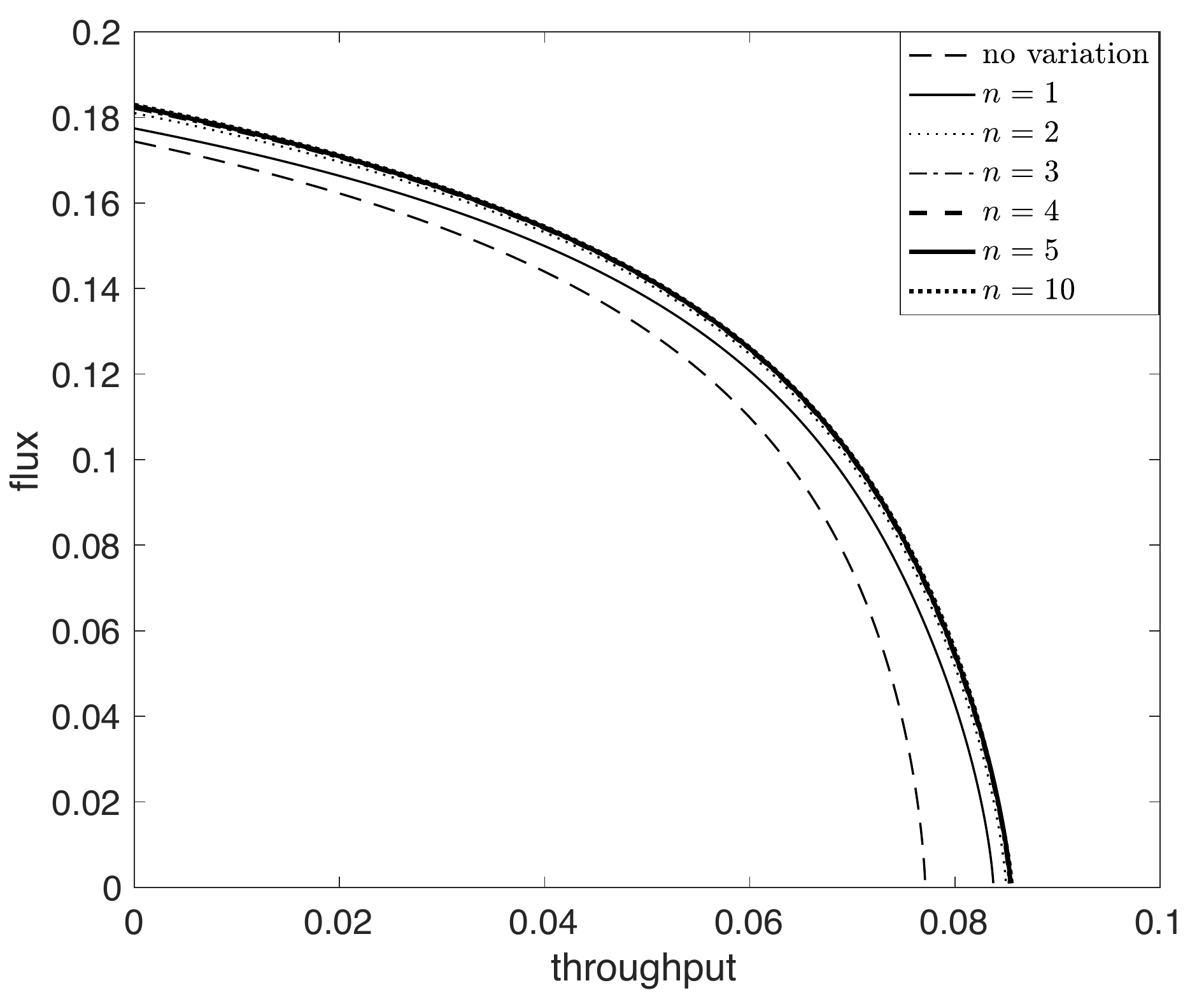}
{\scriptsize (b)}\includegraphics[scale=0.32]{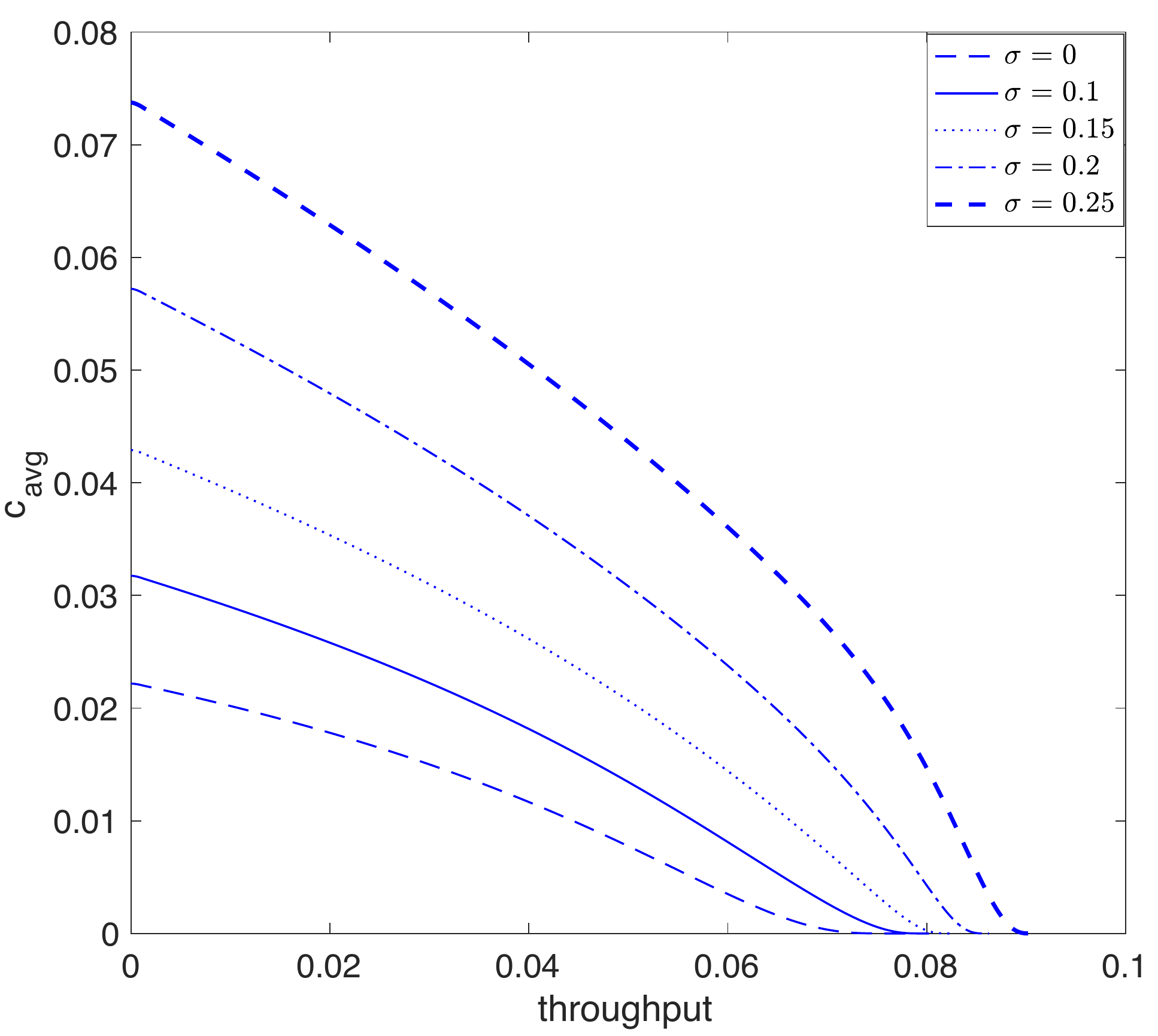}
{\scriptsize (e)}\includegraphics[scale=0.32]{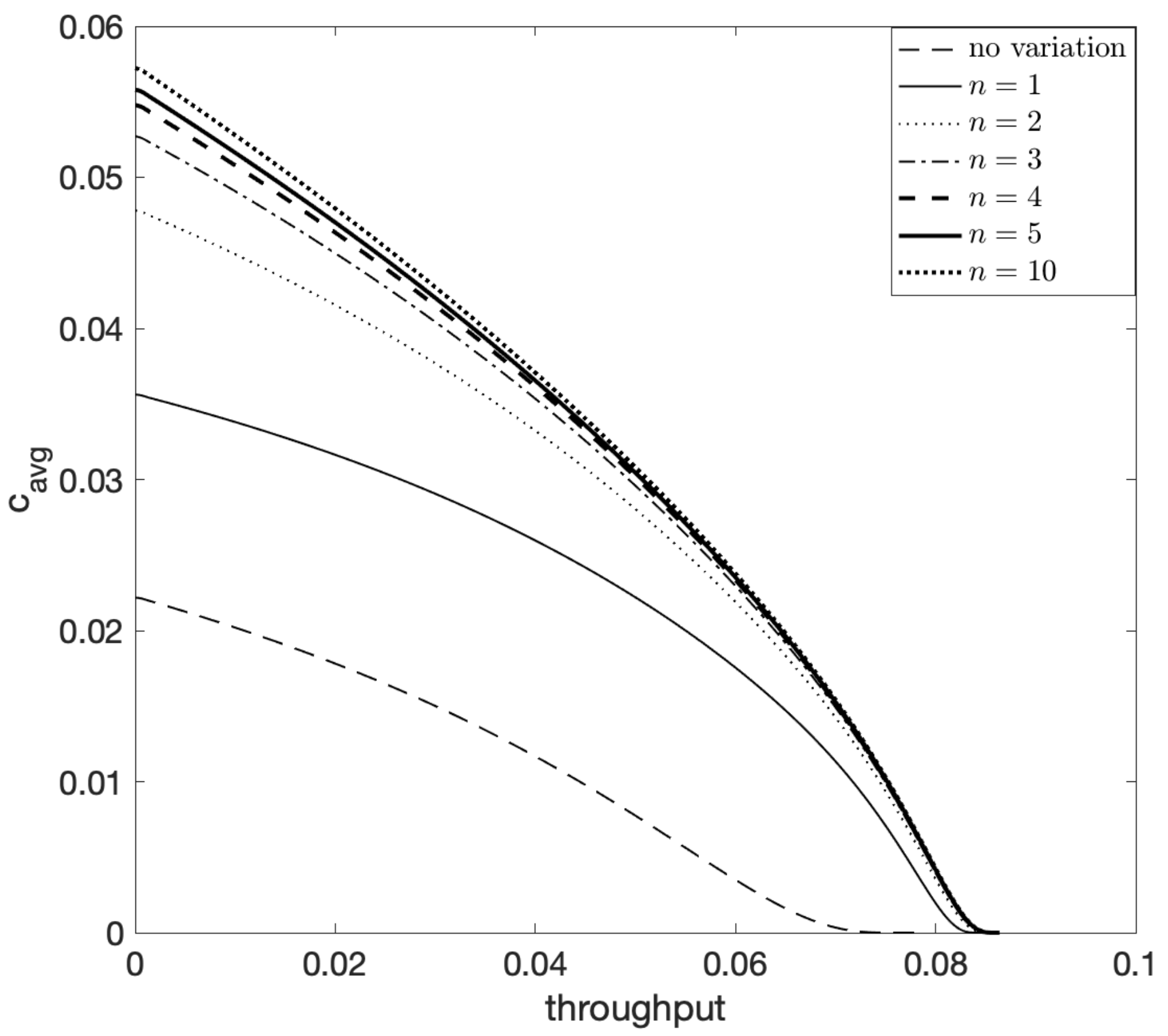}
{\scriptsize (c)}\rotatebox{0}{\includegraphics[scale=.32]{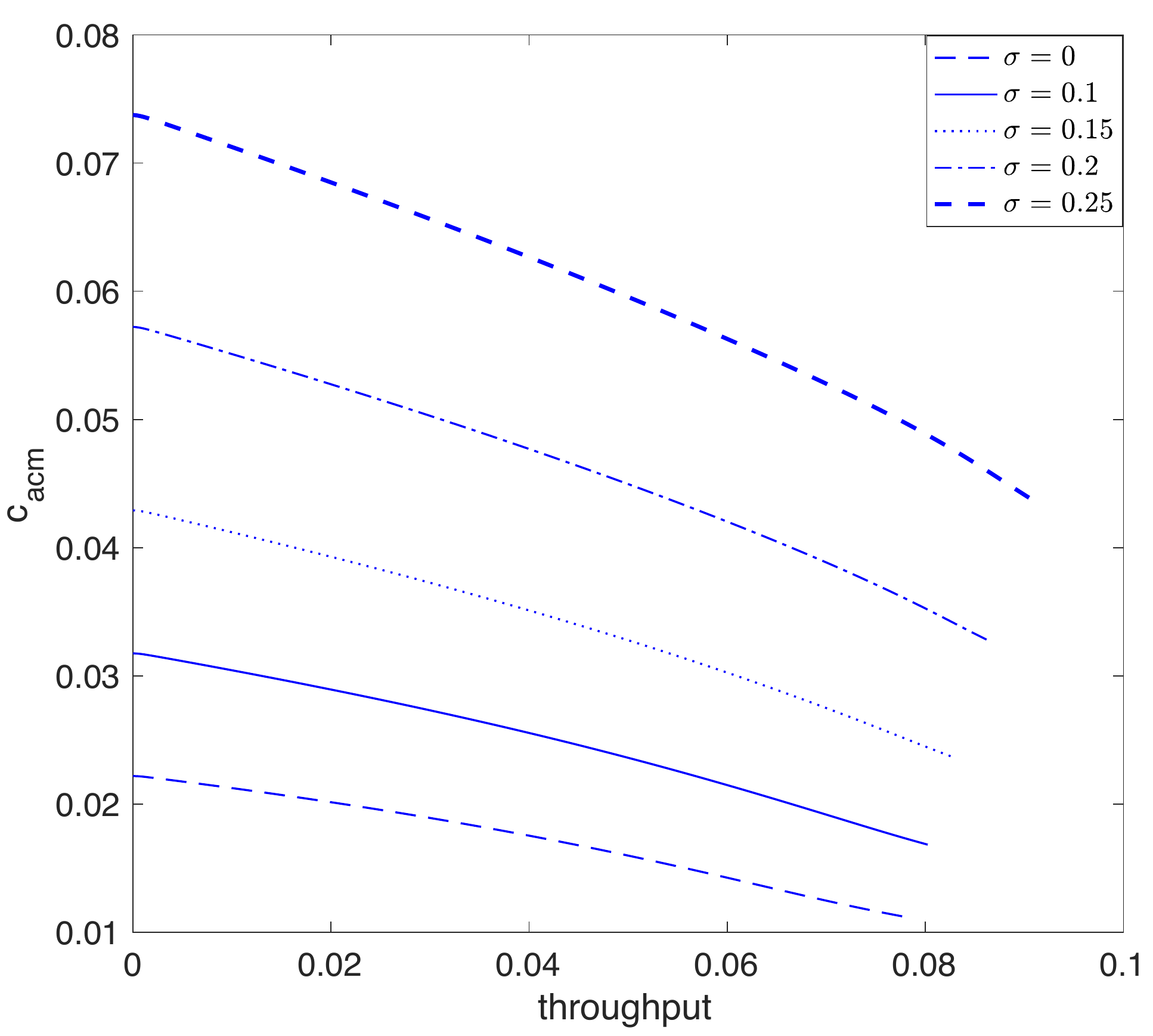}}
{\scriptsize (f)}\includegraphics[scale=0.32]{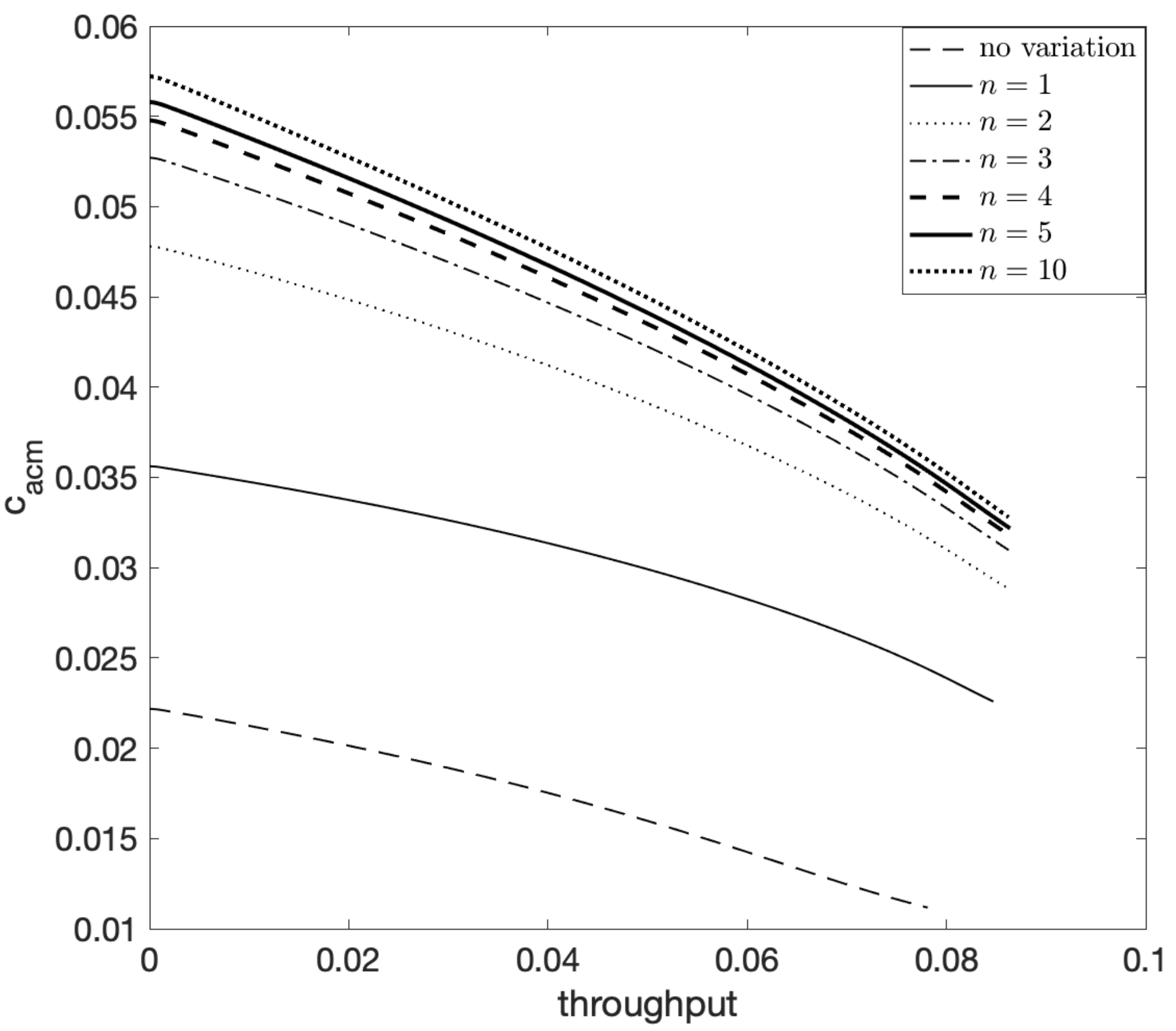}
\caption{\footnotesize{ Effect of in-plane pore-size variation of the form $a(x,\tilde{y},0)= 0.799 (1+\sigma\sin(2n\pi x))$ with $\lambda=1$. (a-c) illustrate variations in $\sigma$ with $n=10$: (a) flux-throughput ( \( \mathcal{F}\)- \( \mathcal{J}\)) plots, (b) $c_{\rm avg}$ (instantaneous average particle concentration in filtrate) vs. throughput \( \mathcal{J}\), and (c)  $c_{\rm acm}$ (particle concentration in accumulated filtrate) vs. throughput  \( \mathcal{J}\). (d-f) illustrate variations in $n$ with $\sigma=0.2$: (d)  \( \mathcal{F}\)- \( \mathcal{J}\), (e)  $c_{\rm avg}$ vs. \( \mathcal{J}\), (f)  $c_{\rm acm}$ vs.  \( \mathcal{J}\). 
}}
\label{2fig_distributional_variation_x}
\end{figure}

Our investigations so far have assumed that pore size does not vary along the length of the pleat, that is, all pores are initially identical (though they evolve differently in time). In practice, however, some variation of pore size is unavoidable in the process of membrane manufacture. This motivates us to explore how variation of the initial pore size distribution in the $x$-direction affects the filtration performance, to gain insight into the importance of membrane homogeneity in applications. 

To study this in a simple and tractable way, we propose the following initial pore profile:
\be
a(x,\tilde{y},0)=(1+\sigma \sin(2n\pi x)) {\rm A}(\tilde{y}), \qquad n=1,2,3,\ldots,
\label{psvar}
\ee
in which $\sigma$ and $n$ capture the amplitude and spatial frequency of variation respectively, and ${\rm A}(\tilde{y})$ is a polynomial in $\tilde{y}$ as considered previously. Note that the $x$-averaged pore size is the same for all cases. Before conducting any optimizations, we first explore variations of the form (\ref{psvar}) in a membrane with simple cylindrical pores, where ${\rm A}$ is
independent of $\tilde{y}$. Figures \ref{2fig_distributional_variation_x}(a,b,c) show the flux-throughput (\( \mathcal{F}\)-\( \mathcal{J}\)) graphs, and plot the average particle concentrations  $c_{\rm avg} $, and the accumulated particle concentrations $c_{\rm acm} $ versus throughput, respectively, as $\sigma$ varies from 0 (uniform pores) to 0.25 (25\% variations in pore sizes) with $n=10$ and $\lambda=1$. We take ${\rm A}=0.799$; this value ensures that the largest pores are (just) contained within the assumed period-box. Increasing the amplitude $\sigma$ of the pore size variation increases total throughput \( \mathcal{J}\)$(t_{\rm f})$; however, the particle concentration in the filtrate increases as well, over the entire duration of filtration. We note that the differences in  \( \mathcal{J}\)$(t_{\rm f})$, though measurable, are not highly significant as $\sigma$ varies ( \( \mathcal{J}(t_{\rm f})\) increases by about 20\% as $\sigma$ increases from 0 to 0.25); however the changes in particle concentration in the filtrate are significant: both $c_{\rm avg}(0)$ and $c_{\rm acm}(0)$ increase by more than 300\% for the same change in $\sigma$.  Figures \ref{2fig_distributional_variation_x}(d,e,f) show the flux-throughput ( \( \mathcal{F}\)-\( \mathcal{J}\)) graphs, and plot the average particle concentrations $c_{\rm avg}$, and the accumulated particle concentrations  $c_{\rm acm}$ versus throughput  \( \mathcal{J}\), respectively, as $n$, the spatial frequency of pore-size variations, increases (with $\sigma=0.2$). The same trends are observed  as for the variation of $\sigma$: there is a modest increase in total throughput  \( \mathcal{J}\)$(t_{\rm f})$ as $n$ increases, and a more significant increase in the particle concentration in the filtrate. We find that for sufficiently large $n$ ($n>10$), its exact value is not relevant. 
Since the increase in \( \mathcal{J}\)$(t_{\rm f})$ is so modest as both $\sigma$ and $n$ increase, while particle retention capability declines significantly, these results  suggest that pore-size variation is unfavorable for effective filtration, and manufacturers should seek to make their membranes as uniform as possible (with respect to in-plane variations).  Future work is needed to test whether this finding extends to more general pore shapes.   

Although uniform pores may well be desirable, some pore-size variation is unavoidable in manufacturing,
\begin{figure}
\centering
{\scriptsize (a)}\rotatebox{0}{\includegraphics[scale=.32]{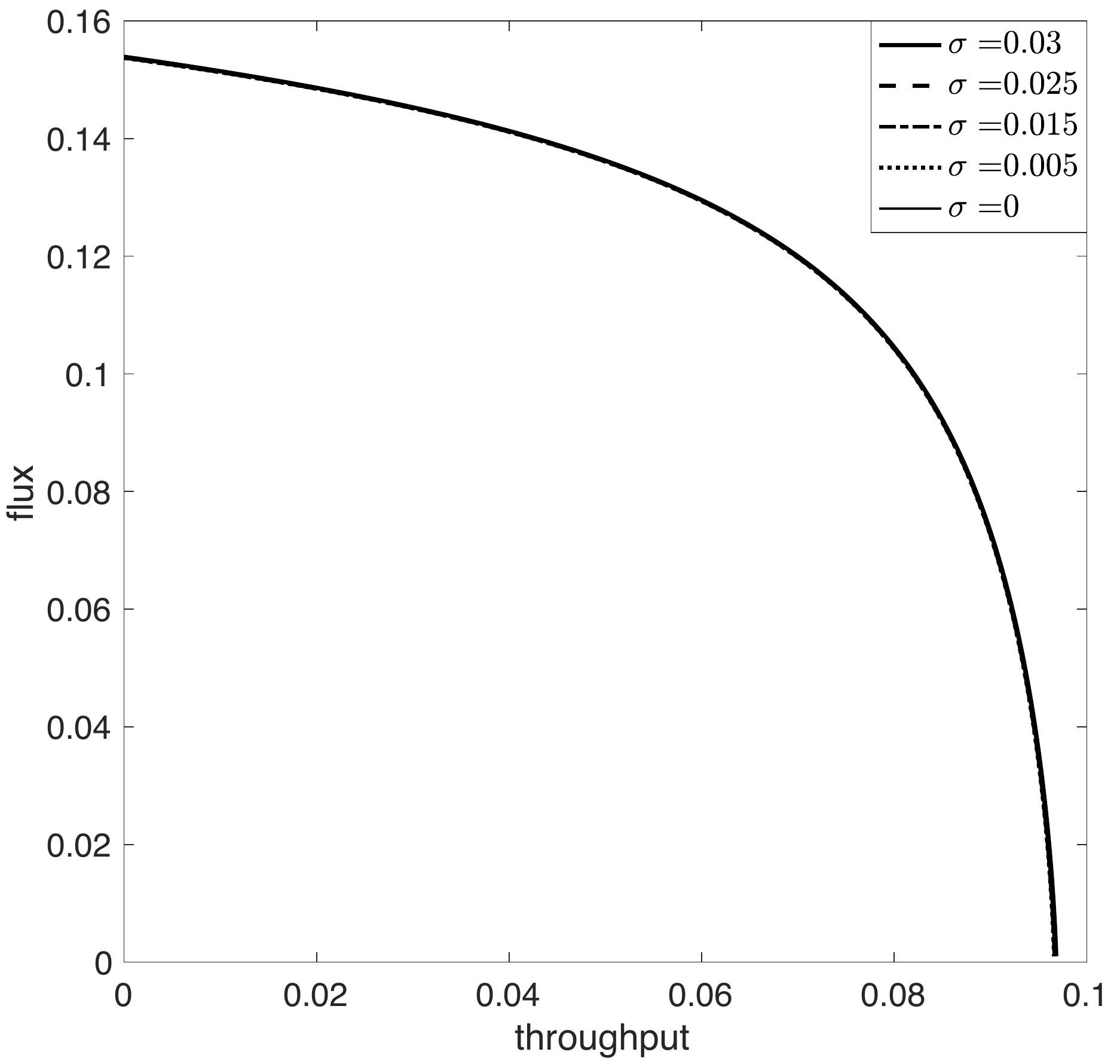}}\\
{\scriptsize (b)}\includegraphics[scale=0.32]{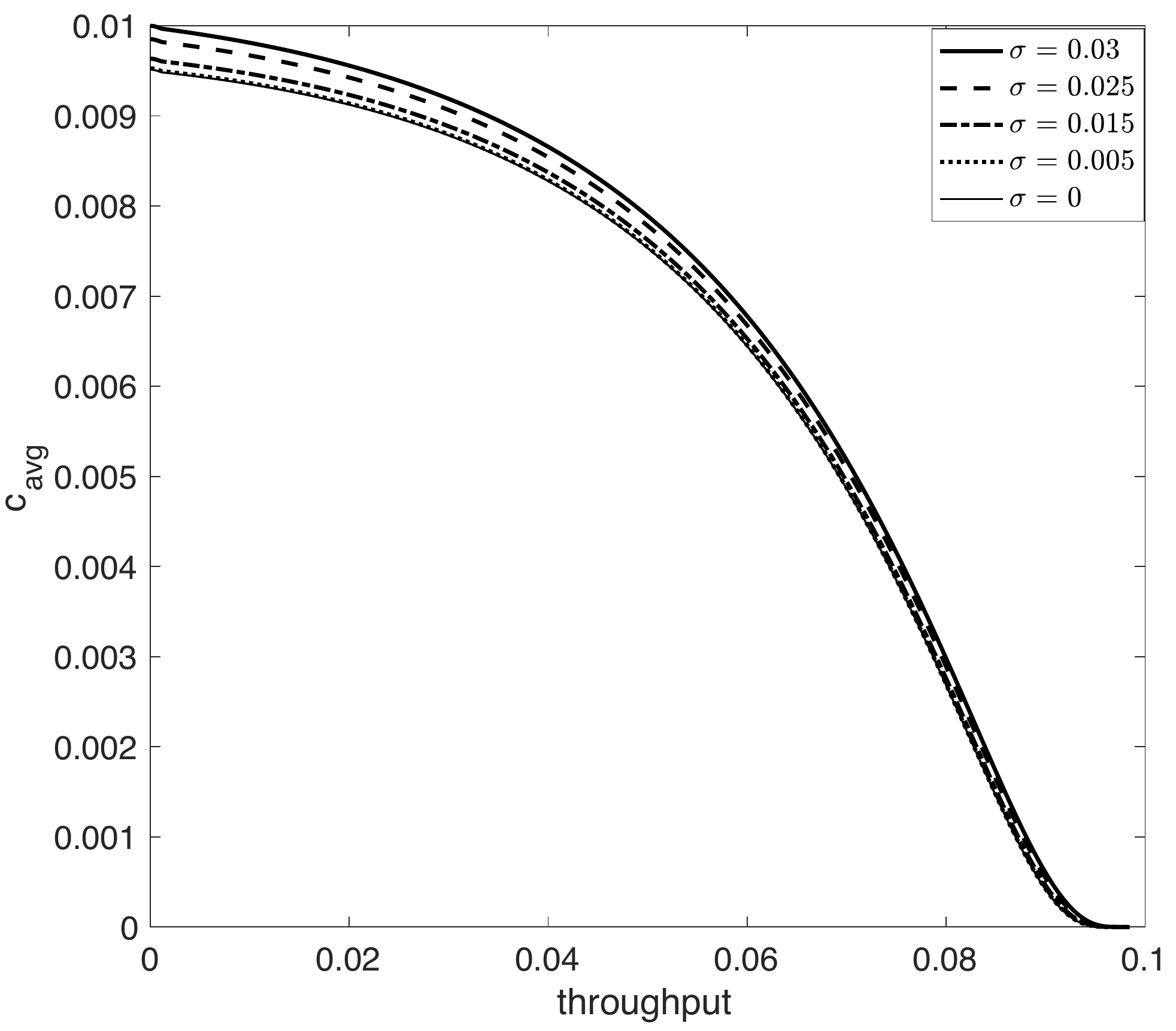}
{\scriptsize (c)}\rotatebox{0}{\includegraphics[scale=.32]{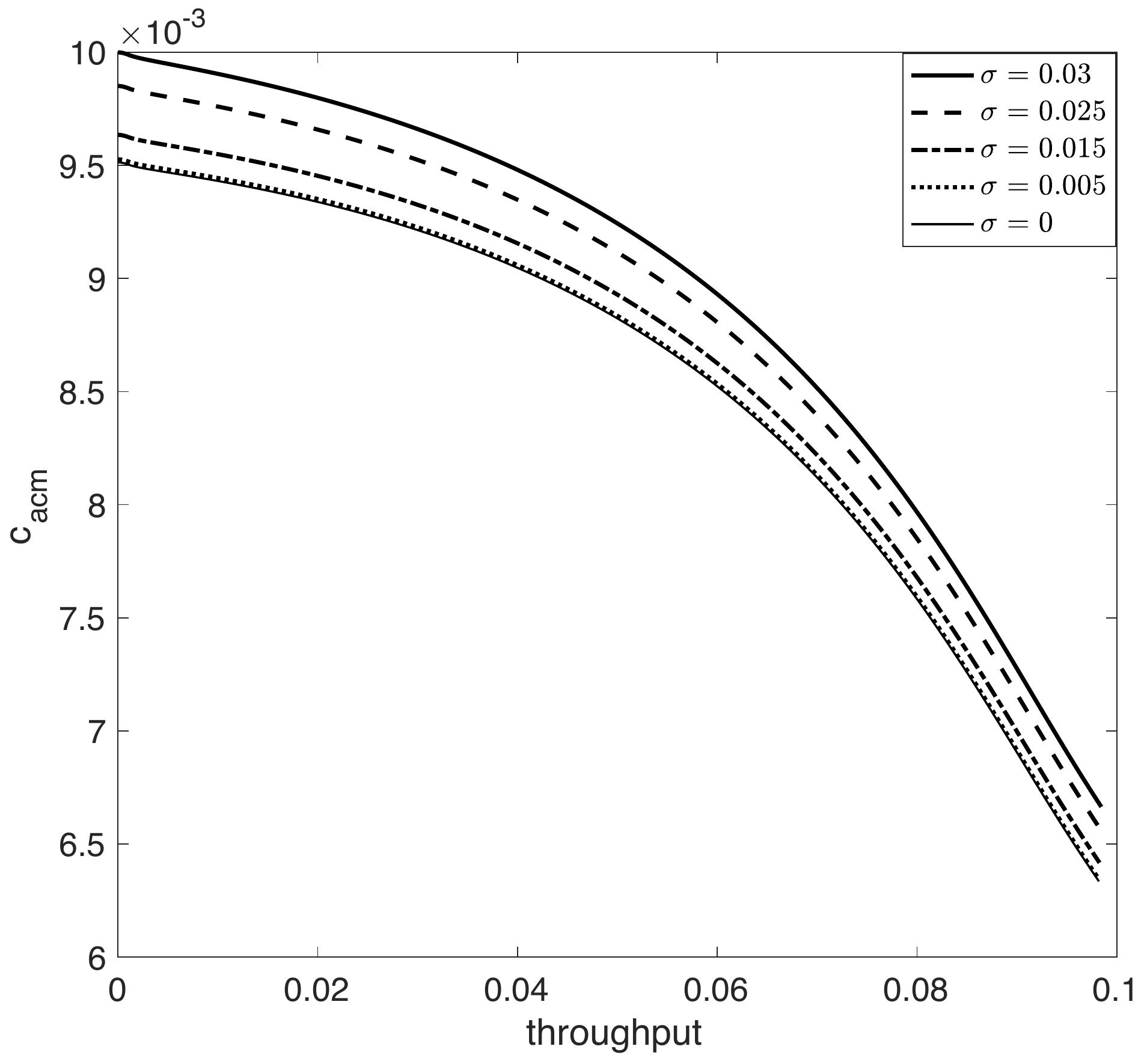}}
\caption{\footnotesize{
Effect of in-plane pore-size variations of the form (\ref{psvar}) $a(x,\tilde{y},0)={\rm A}(\tilde{y})(1+\sigma\sin (2 n\pi x))$.
Results based on this initial pore profile, optimized with $\sigma=0.03$, $n=10$ and ${\rm A}(\tilde{y})$ linear, for $\lambda=1$, $R=99\%$.    
(a) Flux-throughput (\( \mathcal{F}\)-\( \mathcal{J}\)) plots, (b)  $c_{\rm avg}$ vs. throughput \( \mathcal{J}\), and (c)  $c_{\rm acm}$ vs. throughput  \( \mathcal{J}\); for $\sigma=0,0.005,0.015,0.025,0.03$. 
} } 
\label{2fig_optimization_a_0_y_with_variation}
\end{figure}
thus a natural question to ask is: if the maximum pore-size variation is known, what is the optimized pore profile that the manufacturer should aim for? Lacking detailed data on  manufacturing pore-size tolerances, we present some illustrative examples. Suppose that the maximum pore-size variation is $\pm 3\%$. We set $\sigma=0.03$ and $n=10$ initially, in $a(x,\tilde{y},0)$ as given in (\ref{psvar}). With these parameters fixed, we then find the polynomial profiles ${\rm A}(\tilde{y})$ that optimize $a(x,\tilde{y},0)$, using the approach of \S~\ref{2optimization_formulation}. Figure \ref{2fig_optimization_a_0_y_with_variation} shows selected results, optimizing over linear ${\rm A}(\tilde{y})$ only, with the particle removal threshold $R$ fixed at 99\%: in (a,b,c) the {\it dashed} curves represent the {\it optimized} results for  \( \mathcal{F}\)- \( \mathcal{J}\),  $c_{\rm avg}$- \( \mathcal{J}\)and  $c_{\rm acm}$- \( \mathcal{J}\), respectively, carried out for  $\lambda=1$, $n=10$ and $\sigma=0.03$. The remaining curves in (a-c) show results of simulations for this same optimized pore profile ${\rm A}(\tilde{y})$, but taking other values of $\sigma\in (0,0.03)$ in (\ref{psvar}).

\begin{figure}
{\scriptsize (a)}\rotatebox{0}{\includegraphics[width=0.465\linewidth, height=0.39\linewidth]{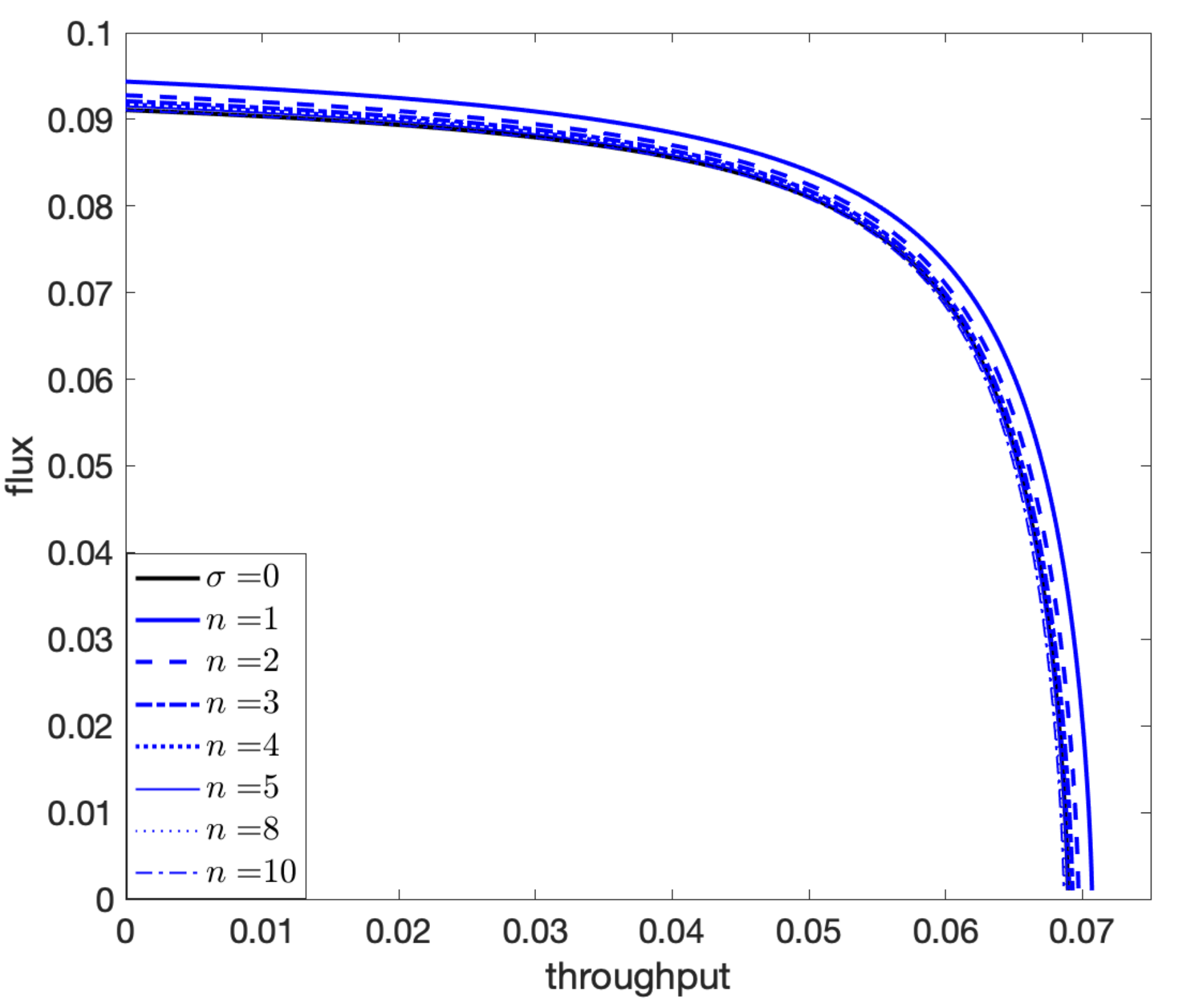}}{\scriptsize (d)}\includegraphics[width=0.465\linewidth, height=0.39\linewidth]{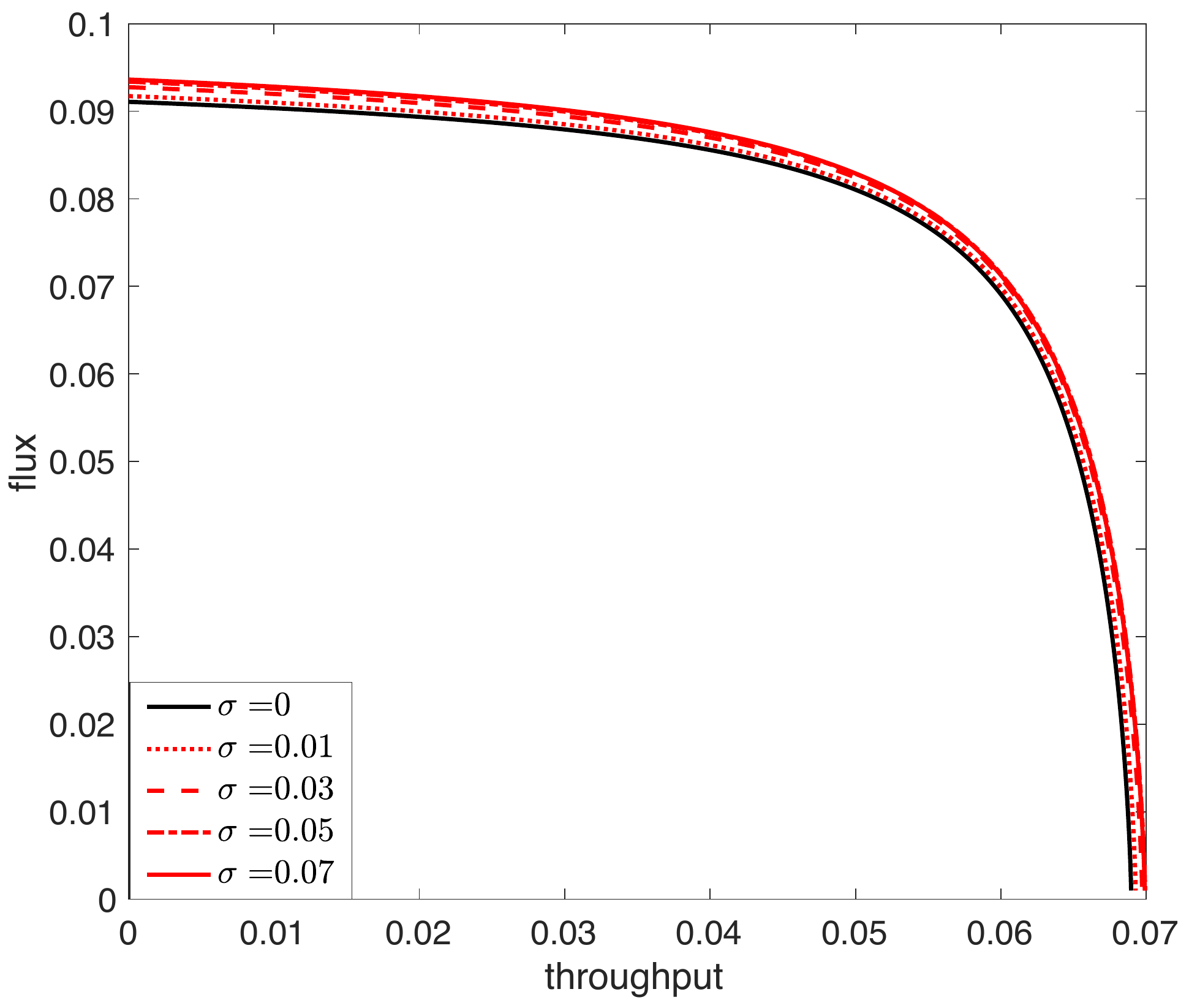}
{\scriptsize (b)}\includegraphics[width=0.465\linewidth, height=0.39\linewidth]{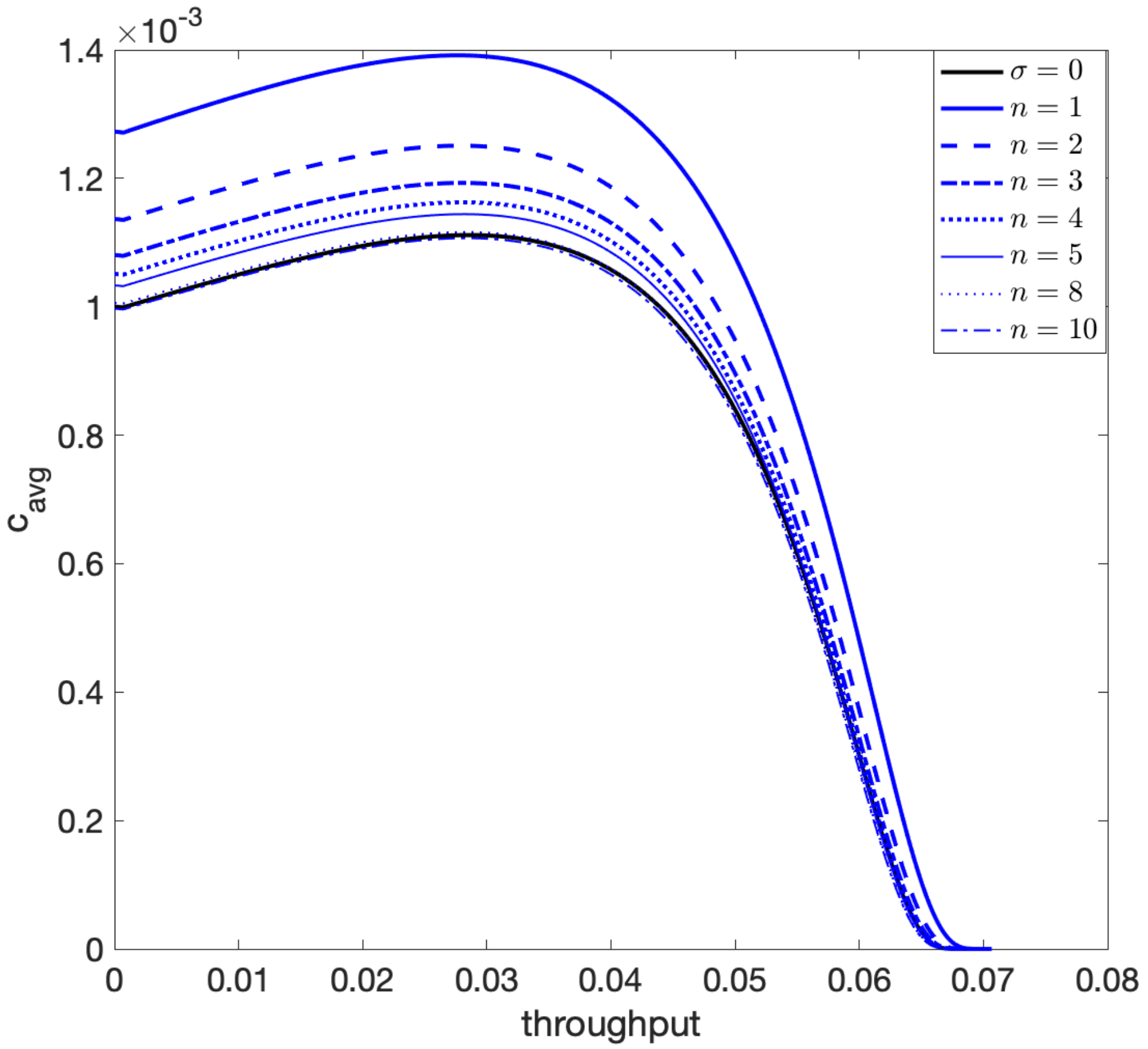}{\scriptsize (e)}\includegraphics[width=0.465\linewidth, height=0.39\linewidth]{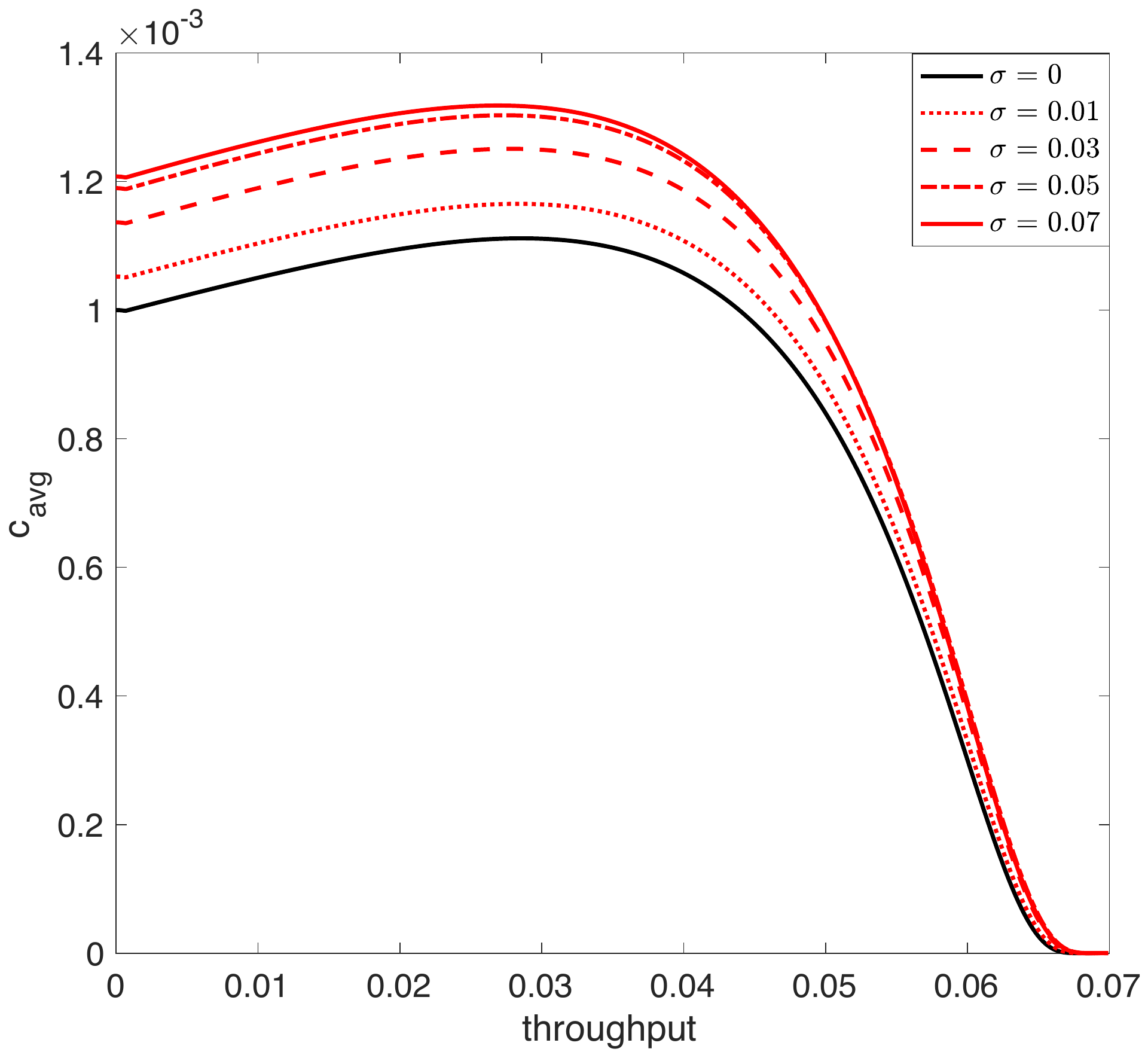}
{\scriptsize (c)}\rotatebox{0}{\includegraphics[width=0.465\linewidth, height=0.39\linewidth]{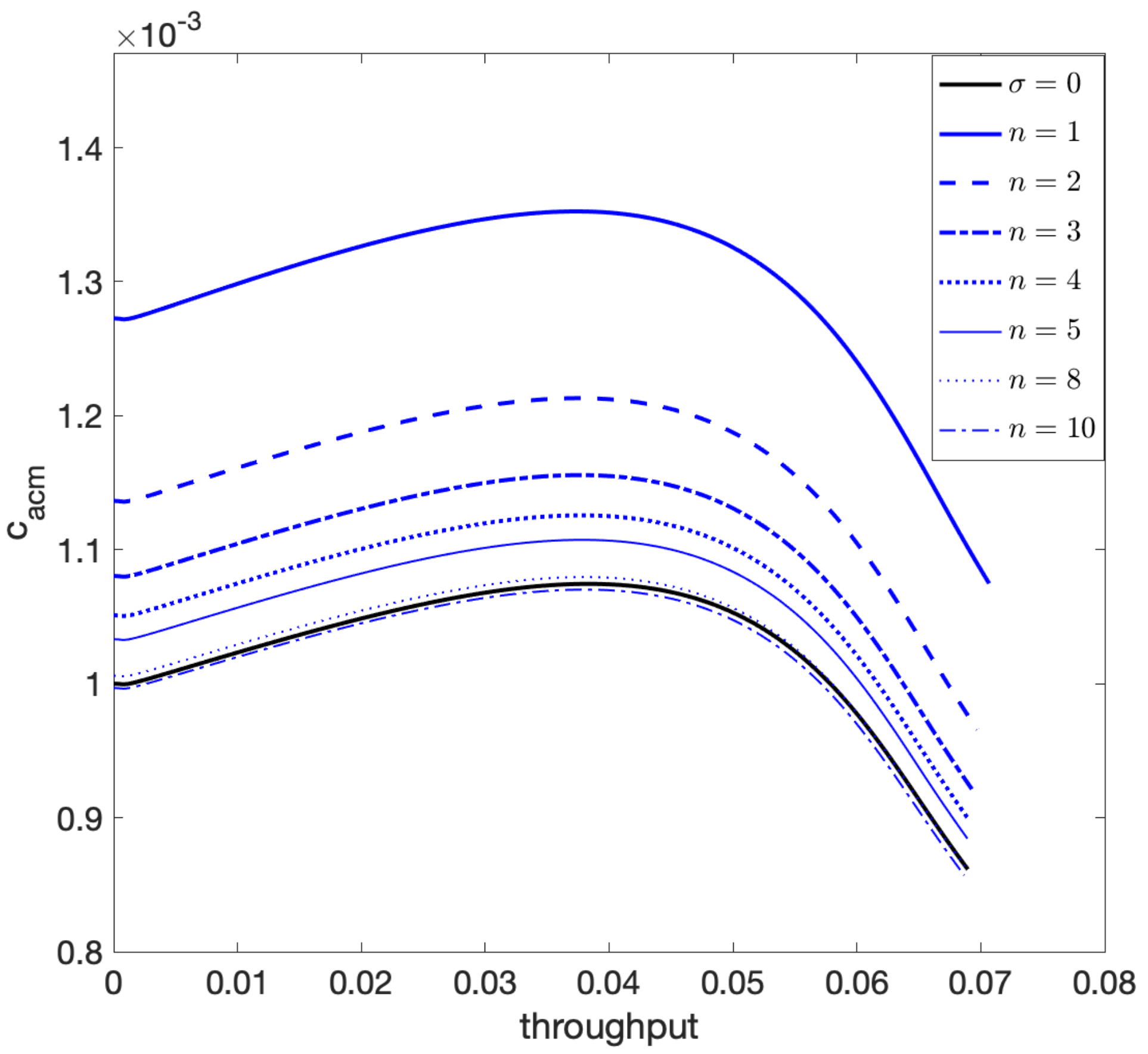}}
{\scriptsize (f)}\includegraphics[width=0.465\linewidth, height=0.39\linewidth]{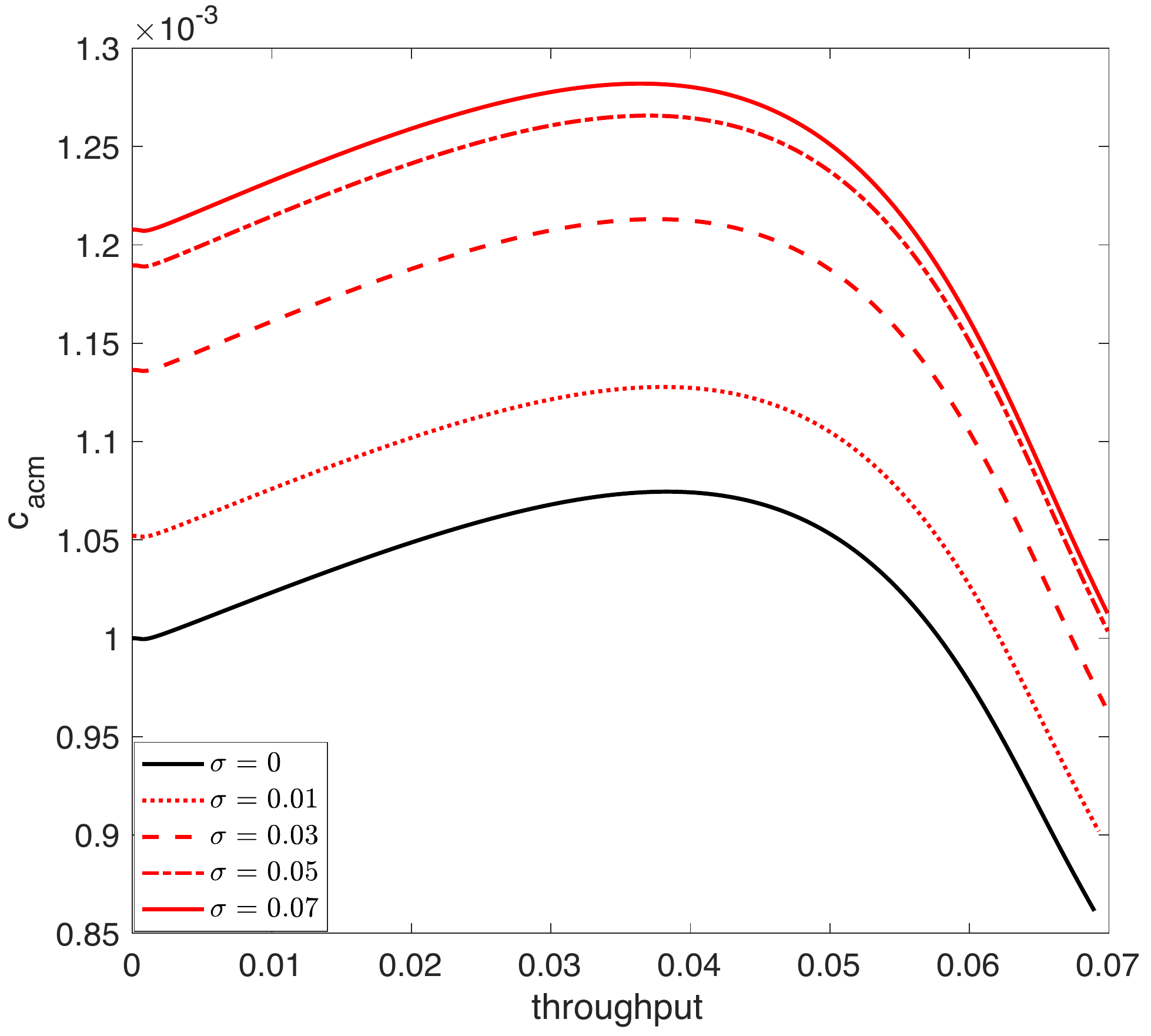}
\caption{\footnotesize{
Effect of in-depth pore-size variations (\ref{psvar_y}) $a(x,\tilde{y},0)={\rm A}(\tilde y)+\sigma \|{\rm A} (\tilde{y}) \|_{L^1}  \frac{\tilde {\rm A}(\tilde y, n) }{\|\tilde {\rm A} \|_{L^\infty} (n)}$:
(a-c) show results with $\sigma=0.03$, and  ${\rm A}(\tilde{y})$ the linear optimized pore profile for $\lambda=1$, $R=99.9\%$.  
(a) Flux-throughput (\( \mathcal{F}\)-\( \mathcal{J}\)) plots, (b)  $c_{\rm avg}$ vs. throughput  \( \mathcal{J}\), (c)  $c_{\rm acm}$ vs. throughput  \( \mathcal{J}\), for $n=1,2,3,4,5,8,10$. 
(d-f) show results with $n=2$: (d)  \( \mathcal{F}\)-\( \mathcal{J}\) plots, (e)  $c_{\rm avg}$ vs. \( \mathcal{J}\), (f)  $c_{\rm acm}$ vs. \( \mathcal{J}\), for $\sigma=0, 0.01, 0.03, 0.05, 0.07$.
} } 
\label{2fig_a_0_y_with_variation_in_y}
\end{figure}

The flux-throughput ( \( \mathcal{F}\)- \( \mathcal{J}\)) plots in figure \ref{2fig_optimization_a_0_y_with_variation}(a) show almost no change as $\sigma$ is varied. Figures \ref{2fig_optimization_a_0_y_with_variation}(b,c) show that, if pore size varies less than the estimated $\pm 3\%$ used for the optimization, the particle removal requirement will not be violated. Both $c_{\rm avg} (t)$ and $c_{\rm acm} (t)$ decrease as $\sigma$ decreases. Figures \ref{2fig_optimization_a_0_y_with_variation}(a-c) collectively indicate that if the manufacturer has a reliable upper bound on the in-plane pore-size variation ($x$-direction), they could use the optimized initial pore profile $a(x,\tilde{y},0)$ (varying in both $x$- and $\tilde{y}$-directions as per (\ref{psvar})) based on this bound. As long as the bound holds, the resulting profile will yield almost the optimized total throughput, without violating the particle removal requirement. Though we here considered optimizing only over the class of linear profiles ${\rm A}(\tilde{y})$, results for quadratic and cubic profiles (not shown here) support the same conclusions.

In addition to in-plane variations in membrane pore size, it is equally inevitable that pores will deviate from the desired design (in both size and shape) in the depth of the membrane ($\tilde{y}$ direction). To gain some initial insight into the effects of such depth-dependent pore profile variation, we consider the following initial pore profile
\be
a(x,\tilde{y},0)={\rm A}(\tilde y)+\sigma \|{\rm A} (\tilde y)  \|_{L^1}  \frac{\tilde {\rm A}(\tilde y, n) }{\|\tilde {\rm A}(\tilde{y},n) \|_{L^\infty}}, \qquad n=1,2,3,\ldots,
\label{psvar_y}
\ee
in which $\sigma$ and $n$ capture the amplitude and spatial frequency of variation, ${\rm A}(\tilde{y})$ is a polynomial in $\tilde{y}$ as considered previously,
and $ \| \cdot \|_{L^1}$, $\| \cdot \|_{L^\infty}$ are the standard $L^1$ and $L^{\infty}$ (maximum) norms (respectively) for $\tilde y \in [-1/2, 1/2]$. While optimization could be carried out for higher-order polynomials, the results we present are for profiles ${\rm A}(\tilde{y})$ linear in $\tilde{y}$. We choose a functional form for $\tilde {\rm A}(\tilde y, n)$ that permits oscillations of specified wavenumber in the depth of the membrane, while preserving pore volume compared with the unperturbed ``optimal'' pore, which mathematically (to leading order in small pore-size perturbations $\sigma$) reduces to the constraint
\bes
\int_{-1/2}^{1/2}  {\rm A}(\tilde y) \tilde {\rm A}(\tilde y, n) d\tilde y=0.
\ees
In the linear case ${\rm A}(\tilde y)=a \tilde y+b$, $\tilde {\rm A}(\tilde y, n) $ can take the form 
\bes
{\rm A}(\tilde{y},n)= \sin(4n \pi \tilde y)+\frac{(4n+1)a \cos((4n+1) \pi \tilde y)}{8nb}.
\ees
The formulation of (\ref{psvar_y}) then incorporates the idea that the value of $\sigma$ captures the percentage of pore-size amplitude variation from the unperturbed (optimized) pore profile ${\rm A}(\tilde y)$ (e.g. $\sigma=0.03$ corresponds to approximately $3\%$ variation from ${\rm A}(\tilde y)$).

Sample results are shown in figure~\ref{2fig_a_0_y_with_variation_in_y}, where the optimization is carried out with $\lambda=1$, and particle removal threshold $R=99.9\%$ (the optimized result is indicated by $\sigma=0$ in the legend).
Similar to our earlier results of figure~\ref{2fig_distributional_variation_x} for in-plane pore-size variation, figures \ref{2fig_a_0_y_with_variation_in_y}(a,b,c) show the flux-throughput (\( \mathcal{F}\)-\( \mathcal{J}\)) graphs for each case, as well as plots of the instantaneous average particle concentrations $c_{\rm avg}$ and the accumulated particle concentrations $c_{\rm acm}$ in the filtrate  versus  \( \mathcal{J}\), respectively, as $n$, the spatial frequency of pore-size variations in depth, increases (with $\sigma=0.03$ fixed for all except the black $\sigma=0$ curves).  We find that, for $n$ sufficiently large ($n>8$), the effect of the variations is insignificant, with filtration performance approaching that of the unperturbed, optimal, case. In all cases, there is little effect on  \( \mathcal{F}\)- \( \mathcal{J}\) characteristics, but small values of $n$ ($n=1,2,3$) can have rather a large effect on the particle retention capability, leading to as much as 30\% more particles evading capture by the membrane.

Figures~\ref{2fig_a_0_y_with_variation_in_y}(d,e,f) plot the corresponding graphs as the perturbation amplitude $\sigma$ varies from 0 (no variation) to 0.07 (7\% variations in pore sizes) with perturbation wavenumber fixed at $n=2$.  From figure  \ref{2fig_a_0_y_with_variation_in_y}(d) we again observe negligible changes to the flux and throughput characteristics. Particle removal capability, however, changes by nearly $20\%$ for a $5\%$ perturbation of pore radius  indicated by figure  \ref{2fig_a_0_y_with_variation_in_y}(e) and (f). This suggests that particle removal capability is more sensitive to perturbations of (optimized) pore profiles in the $\tilde{y}$ direction compared to in-plane variations  (see figure \ref{2fig_optimization_a_0_y_with_variation} (b) and (c)), and maintaining the optimized pore shape (or more generally, the desired depth permeability gradient of the filter) will be critical to achieve the maximum total throughput while simultaneously satisfying the particle removal requirement.

\section{Conclusion\label{2discussion}}

{\color{black}
We have formulated a simple mathematical model for evaluating the performance of a pleated membrane filter, with variable internal pore structure within the membrane. In order to obtain a model that is fast to simulate, while still capturing the permeability gradients that exist in real filtration membranes, we assume that the membrane pores are tubes of circular cross-section spanning the membrane from upstream to downstream side, and that the variation in the pore radius models the variations in permeability. Though clearly an oversimplification for many membranes, this type of geometrical pore model is frequently used in filtration modeling, and we expect our results to provide a good guide as to how average pore size should vary in the depth of the membrane.

The simplicity of our model allows for quick simulations of filtration all the way to final pore-blocking, which in turn allows us to carry out the optimization of the filter pore profile for a common filtration objective (maximizing throughput over the filter lifetime) and operating condition (a specified partical removal constraint must be satisfied); see \S~3.2. For this filtration objective and operating condition, we are able to use our model to find (numerically) the optimal initial pore shape within a restricted class of such shapes (which, in the interest of keeping computational time relatively short, we take to be low-order polynomials in the depth of the membrane). Our results indicate that this optimization should be sufficient for most practical applications: as the degree of the polynomial increases from 1 to 3 convergence of the results (presumably to some global optimum) appears rapid. We are currently working on possible approaches to solve efficiently the general optimization problem in order to determine the optimum initial pore profile over all possible shapes. 

In \S~\ref{variability} we also briefly explored the impact of (unavoidable) in-plane variations and in-depth variations to the desired (optimal) pore geometry, on filtration performance. Our initial investigations indicate that, for in-plane variations our  optimization techniques could still be useful if the tolerance in pore-size variation is well-characterized, and sufficiently small.  However our investigation for in-depth pore-size variation indicates that particle removal can be significantly impacted by variations, and maintaining the optimized pore shape will be critical to achieve the highest total throughput  \( \mathcal{J}\)$(t_{\rm f})$ while simultaneously satisfying the particle removal requirement. 

One significant observation from our simulations is that, under certain conditions, the particle concentration in the filtrate may {\it increase} after the filtration starts --- that is, particle removal capability of the membrane may actually deteriorate in the early stages of the filtration. This phenomenon, which is known to occur experimentally~\citep{jackson2014}, has not (to the best of our knowledge) been observed in earlier theoretical studies of adsorptive fouling. However, the results of \cite{jackson2014} in fact show monotone deterioration in particle retention, unlike our results which indicate an eventual improvement in retention as significant fouling occurs. This difference indicates that additional refinements to our model may be needed.

Although we believe that our model represents a valuable step forward in helping manufacturers identify optimal membrane structures for given filtration scenarios, it does have several limitations, and there are many potential areas for improvement. First, we only consider very simple homogeneous feed solutions that contain identical particles. In most applications there will likely be multiple species in the same feed, and the objective with respect to which we optimize could also be more complicated: for example, to remove some species while allowing others to pass through. Future work will address filtration with more complex feed solutions.
Second, in this work we only consider one fouling mechanism, while in practice there could be multiple simultaneous mechanisms. It would not be difficult to include multiple fouling mechanisms in our model; for example following the approach of \cite{sanaei2016}. We note, however, that the more mechanisms we include in our model, the more unknown parameters the model will have and the larger the parameter space to explore. For applications where standard blocking dominates, our model presented here should be adequate. Third, even for such applications where standard blocking dominates, there are further details that could and should be considered. For example, particles deposited on the clean membrane in the initial stages of filtration could have a shielding effect, and modify the physico-chemical interactions between particles and membrane, making it possibly more difficult for particles arriving at a later time to deposit on the membrane.
Again, this is something we plan to address in our future work.
Lastly, though we believe our simple ``tubular pore'' model should provide a good guide as to desirable membrane properties, there are certainly other types of pore structures that could be considered, e.g. branching pore structures. Some preliminary studies may be found in \citet{sanaei2018}, and we are currently undertaking more ambitious studies into how membranes with arbitrary pore networks can be efficiently modeled.}

\subsection*{Acknowledgements} 

Y.S. thanks Thilo Simon for several helpful discussions which led to Appendix~\ref{symmetry}. All authors acknowledge financial support from the National Science Foundation under
Grants NSF-DMS-1261596 and NSF-DMS-1615719. P.S. was supported in part by the NSF Research
Training Group in Modeling and Simulation Grant RTG/DMS-1646339.

\appendix

\section{Symmetry in the pore profile evolution\label{symmetry}}

In our simulations we observed that if the initial pore profile distribution is symmetric about $x=1/2$ (i.e. $a_0(x,\tilde{y})=a_0(1-x,\tilde{y})$),  the pores will retain this symmetry for the entire filtration process (i.e. $a(x,\tilde{y}, t)=a(1-x,\tilde{y}, t)$, $0\leq t\leq t_{\rm f}$), and we here present a proof for this. This symmetry arises from the coupled ODEs (\ref{p0+xx_s}) and (\ref{p0-xx_s}), reproduced below: 
\be 
{p_0}^+_{xx}(x, t)=\frac{\Gamma (p_0^+(x, t)-p_0^-(x, t))}{r_{\rm m}(x,t)},
\quad p_0^+(0, t)=1,\quad {p_0}^+_{x}(1, t)=0,
\label{p0+xx_sym_A} \\
-{p_0}^-_{xx}(x,t)=\frac{\Gamma (p_0^+(x, t)-p_0^-(x, t))}{r_{\rm m}(x,t)},\quad p_0^-(1, t)=0,\quad {p_0}^-_{x}(0, t)=0,
\label{p0-xx_sym_A}
\ee
(recall that the net membrane resistance $ r_{\rm m}(x, t)=\int_{ -1/2 }^{1/2 }{ {a^{-4}(x,\tilde{y},t)} d \tilde{y}}$). Observe first that, showing the pore profile evolution is symmetric about $x=1/2$ given the initial pore profile is symmetric about $x=1/2$, is equivalent to showing that statement (\ref{root}) holds:
\be
r_{\rm m}(x, t)=r_{\rm m}(1-x, t)  \quad \implies \quad p_0^+(x, t)-p_0^-(x, t)=p_0^+(1-x, t)-p_0^-(1-x, t).~~~
\label{root}
\ee
Note that 
\[
a(x,\tilde{y}, t)=a(1-x,\tilde{y}, t) \quad  \implies  \quad
r_{\rm m}(x, t)=r_{\rm m}(1-x, t).
\]
At $t=0$, we have
\[
a(x,\tilde{y},0)=a_0(x,\tilde{y})=
a_0(1-x,\tilde{y})=a(1-x,\tilde{y}, 0) \quad \implies \quad 
r_{\rm m}(x, 0)=r_{\rm m}(1-x, 0), 
\]
and $ c(x,1/2, 0)=c(1-x,1/2, 0)=1$ (in fact, $c(x,1/2, t)=c(1-x,1/2, t)=1$ for $0\leq t\leq t_{\rm f}$). From equation (\ref{deposition_nd_asmp_s}), reproduced below:
\begin{equation}
\frac{\partial c (x,\tilde{y},t)}{\partial \tilde{y}}=\frac{\lambda a(x,\tilde{y},t) c(x,\tilde{y},t) r_{\rm m}(x,t)}{p_0^+(x, t)-p_0^-(x, t)} ,\quad c(x,\frac{1}{2} ,t)=1,
\label{deposition_nd_asmp_s_a}
\end{equation}
we see that if $p_0^+(x, 0)-p_0^-(x, 0)=p_0^+(1-x, 0)-p_0^-(1-x, 0)$, then we have $c (x,\tilde{y},0)=c (1-x,\tilde{y},0)$.  From equation (\ref{shrinkage_nd_s}), also reproduced below:
\begin{equation}
\frac{\partial a(x,\tilde{y},t)}{\partial t}= - c(x,\tilde{y},t), \quad a(x,\tilde{y},0)=a_0 (\tilde{y}),
\label{shrinkage_nd_s_a}
\end{equation}
we then know that $a(x,\tilde{y}, \Delta t)=a(1-x,\tilde{y}, \Delta t)$ for the next time step $t=0+\Delta t$. Thus, the evolution will not break the symmetry as long as we have $p_0^+(x, t)-p_0^-(x, t)=p_0^+(1-x, t)-p_0^-(1-x, t)$ for each $t$. Hence, in the following we suppress the time dependence and focus on showing that (\ref{root}) holds. 

\begin{proof}

To show a function $f(x)$ is symmetric about $x=1/2$, we just need to show $f(x)=f(1-x)$. In our case, $f(x)=p_0^+(x)-p_0^-(x)$.

 Let $k(x)=\Gamma/r_{\rm m}(x) >0$, where we know from the above that $k(x)$ is also symmetric about $x=1/2$ if $r_{\rm m}$ is, i.e. $k(x)=k(1-x)$. The coupled ODEs (\ref{p0+xx_sym_A}) and (\ref{p0-xx_sym_A}) can then be written in the following form (here we use an overdot to denote $d/dx$):
\be
\ddot p_0^+=k(p_0^+-p_0^-), \label{ddx1}\\
\ddot p_0^-=-k(p_0^+-p_0^-),  \label{ddx3}\\
\quad p_0^+(0)=1, \quad \dot p_0^+(1)=0,
\quad p_0^-(1)=0, \quad \dot p_0^-(0)=0. \label{bcx1x3}
\ee
From (\ref{ddx1}) and (\ref{ddx3}), we obtain the following:
\be
p_0^+=-p_0^-+Ax+B,\label{x1intermsofx3}
\ee
where $A$ and $B$ are just arbitrary constants. From (\ref{x1intermsofx3}) and (\ref{bcx1x3}), we can get another four boundary conditions in terms of $A$ and $B$:
\bes
p_0^+(1)=A+B,\\
\dot p_0^+(0)=A,\\
p_0^-(0)=B-1,\\
\dot p_0^-(1)=A.
\ees
Now let $w(x)=[p_0^+(x)-p_0^-(x)]-[p_0^+(1-x)-p_0^-(1-x)]$. We claim that
\begin{equation*}
w(x)=0, \forall x \in[0,1].
\end{equation*}
Checking the boundary conditions for $w(x)$ we get 
\bes
w(0)=A+2B-2,\\
w(1)=-A-2B+2,\\
\dot w(0)=A-A=0,\\
\dot w(1)=-A+A=0.
\ees
From the coupled ODEs (\ref{ddx1}), (\ref{ddx3}), (\ref{bcx1x3}) and the symmetric condition of $k(x)$, 
$w(x)$ must satisfy the following system: 
\bes
\ddot w(x)=2k(x)w(x),\\
\dot w(0)=0,\\
\dot w(1)=0.
\ees
Clearly, $w(x)\equiv0$ is a solution. To show the uniqueness of this solution, we use an energy type argument. 
Consider the following:
\bes
\ddot {w^2}=2\dot w^2+2w\ddot w=2\dot w^2+4k(x)w^2,
\ees where  $\ddot {w^2}=\frac{d^2 (w^2)}{dx^2}$. 
Assume $w^2(\ge0)$ takes a maximum at $x_0\in(0,1)$, then $\ddot {w^2}(x_0)\le0$, which implies
\bes
0\le2\dot w^2(x_0)+4k(x_0)w^2(x_0)=\ddot {w^2}(x_0)\le0,\\
\implies
w^2(x_0)=0, \quad \dot w^2(x_0)=0, \\
\implies
\text{max}~ w^2(x)=0,  \forall x \in(0,1).
\ees
Since $w(x)\equiv0$ is a solution, we know $w(x)=0,  \forall x \in(0,1).$ Now we just need to check the boundary to make sure $w(0)=w(1)=0$. Assume $w^2(0)=C>0$, we know $\dot {w^2}(0)=0$, and $\ddot {w^2}(x)\ge0, \forall x \in[0,1]$. By Taylor's Theorem, for some $x_0\in (0,1)$ 
 we have
\bes
w^2(x_0)=w^2(0)+\frac{\ddot {w^2}(\xi)}{2}x^2 \ge C>0 ~\text{for some} ~\xi \in (0, x_0)~ \# 
\ees
which implies $w(0)=0$. Similarly, we can show $w(1)=0$. We have thus proved the symmetry since we have shown
\bes
w(x)\equiv0, ~\forall x \in[0,1].
\ees

\end{proof}



\bibliography{filtration}   

\end{document}